\documentclass[%
 aip,
 reprint,%
 ]{revtex4-1}
\usepackage[utf8]{inputenc}
\usepackage[T1]{fontenc}

\expandafter\let\csname equation*\endcsname=\relax
\expandafter\let\csname endequation*\endcsname=\relax
\usepackage{amsmath,amssymb,amsthm,amsbsy} 
\usepackage{braket}
\usepackage{orcidlink}
\usepackage[final]{microtype}
\usepackage{mathtools}
\mathtoolsset{mathic=true}
\usepackage{graphicx}
\usepackage{physics}
\usepackage[capitalize]{cleveref}
\crefformat{appendix}{#2#1#3}
\usepackage{bm}
\usepackage{todonotes}

\usepackage{changes}
\usepackage{float}
\usepackage{acronym}
\usepackage{placeins}
\usepackage{enumitem}
\setlist[description,1]{leftmargin=0.4cm, itemindent=-.2cm}

\graphicspath{{./figures}}

\def\sys{\ensuremath{\mathrm{S}}}
\def\bath{\ensuremath{\mathrm{B}}}
\def\inter{\ensuremath{\mathrm{I}}}
\def\nth{\ensuremath{^{(n)}}}

\newcommand{\mat}[1]{\ensuremath{{\underline{{#1}}}}}
\def\kmat{{\mat{k}}}

\newcommand{\cc}{\ensuremath{\mathrm{c.c.}}}
\newcommand{\hc}{\ensuremath{\mathrm{h.c.}}}
\def\iu{\ensuremath{\mathrm{i}}}
\def\i{\iu}
\newcommand*{\id}{\text{\usefont{U}{bbold}{m}{n}1}}

\def\dim{\ensuremath{\operatorname{dim}}}
\def\hilb{\ensuremath{\mathcal{H}}}

\def\eu{\ensuremath{\operatorname{e}}}

\newcommand{\refcite}[1]{Ref.\ \cite{#1}}

\makeatletter
\newcommand*{\rom}[1]{\expandafter\@slowromancap\romannumeral #1@}
\makeatother
\newcommand{\phase}[1]{\textbf{\rom{#1}}}

\DeclareMathOperator{\bosedist}{\overline{n}}
\DeclareDocumentCommand\bose{}{\opbraces{\bosedist}}

\def\cold{\ensuremath{\mathrm{cold}}}
\def\hot{\ensuremath{\mathrm{hot}}}

\newacro{HEOM}{Hierarchical Equations of Motion}
\newacro{HOPS}{Hierarchy of Pure States}
\newacro{NMQSD}{Non-Markovian Quantum State Diffusion}
\newacro{QED}{Quantum Electrodynamics}
\newacro{MCTDH}{Multi-Configuration Time-Dependent Hartree}
\newacro{TEMPO}{Time-Evolving Matrix Product Operator}
\newacro{OLC}{Otto-like cycle}
\newacro{BCF}{bath correlation function}
\newacro{SD}{spectral density}

\makeatletter
\def\@email#1#2{%
 \endgroup
 \patchcmd{\titleblock@produce}
  {\frontmatter@RRAPformat}
  {\frontmatter@RRAPformat{\produce@RRAP{*#1\href{mailto:#2}{#2}}}\frontmatter@RRAPformat}
  {}{}
}%
\makeatother

\begin{document}

\title{Dynamics of a strongly coupled quantum heat engine -- computing bath observables from the hierarchy of pure states}

\date{\today}



\newcommand{\fixme}[1]{\marginpar{\tiny\textcolor{red}{#1}}}

\author{Valentin Boettcher\,\orcidlink{0000-0003-2361-7874}}
\address{Institute of Theoretical Physics, TU Dresden University of Technology, 01062 Dresden, Germany}
\address{Department of Physics, McGill University, Montréal, Québec, Canada H3A 2T8}

\author{Richard Hartmann\,\orcidlink{0000-0002-8967-6183}}
\address{Institute of Theoretical Physics, TU Dresden University of Technology, 01062 Dresden, Germany}

\author{Konstantin Beyer\,\orcidlink{0000-0002-1864-4520}}
\address{Institute of Theoretical Physics, TU Dresden University of Technology, 01062 Dresden, Germany}
\address{Department of Physics, Stevens Institute of Technology, Hoboken, New Jersey 07030, USA}

\author{Walter T. Strunz\(^{*}\)\orcidlink{0000-0002-7806-3525}}
\address{Institute of Theoretical Physics, TU Dresden University of Technology, 01062 Dresden, Germany} 
\email{walter.strunz@tu-dresden.de}





\begin{abstract}
  We present a fully quantum dynamical treatment of a quantum heat engine \emph{and} its baths based on the Hierarchy of Pure States (HOPS), an exact and general method for open quantum system dynamics. We show how the change of the bath energy and the interaction energy can be determined within HOPS, for arbitrary coupling strength and smooth time dependence of the modulation protocol. The dynamics of all energetic contributions during the operation can be carefully examined both, in its initial transient phase and also later, in its periodic steady state. A quantum Otto engine with a qubit as inherently nonlinear work medium is studied in a regime where the energy associated with the interaction Hamiltonian plays an important role for the global energy balance and, thus, must not be neglected when calculating its power and efficiency.
 We confirm that the work required to drive the coupling with the baths depends sensitively on the speed of the modulation protocol. Remarkably, departing from the conventional scheme of well-separated phases by allowing for temporal overlap, we discover that one can even gain energy from the modulation of the bath interactions. 
We visualize these various work contributions using the analogue of state change diagrams of thermodynamic cycles.
  We offer a concise, full presentation of HOPS with its extension to bath observables, as it serves as a universal tool for the numerically exact description of general quantum dynamical (thermodynamic) scenarios far from the weak-coupling limit. 
\end{abstract}

\maketitle

\section{Introduction}
Quantum thermodynamics tries to extend the powerful classical theory to the realm where the dynamics of the involved systems cannot be described by classical physics \cite{Binder2018}.
First models and results have been investigated way back in the last century~\cite{alickiQuantumOpenSystem1979, Geva1992Feb}.
Through the enormous experimental advances in preparing, controlling, and measuring small individual quantum systems during the last decades, however, the subject has gained significant attention more recently.
Quantum thermodynamic setups are no longer limited to thought experiments, but can actually be implemented on a broad variety of platforms such as trapped ions~\cite{anExperimentalTestQuantum2015a}, superconducting circuits~\cite{zhangExperimentalDemonstrationWork2018,arracheaEnergyDynamicsHeat2023}, semiconductor quantum dots~\cite{josefssonOptimalPowerEfficiency2019,jalielExperimentalRealizationQuantum2019}, or nuclear spins~\cite{batalhaoExperimentalReconstructionWork2014a,petersonExperimentalCharacterizationSpin2019,palExperimentalDemonstrationValidity2019,micadeiExperimentalValidationFully2021,lisboaExperimentalInvestigationQuantum2022,joshiExperimentalInvestigationQuantum2022,klatzowExperimentalDemonstrationQuantum2019,jiSpinQuantumHeat2022}.

Among the various quantum thermodynamic scenarios that have been proposed and investigated, quantum heat engines have always been of particular interest.
It is the conversion of heat into work which made classical thermodynamics a practically very relevant theory and an extension to the quantum realm seems natural. However, it turns out that this task is not as straightforward as one might expect. The concepts of work and heat have proved to be elusive in a quantum setting, and there is an ongoing debate on a meaningful (measurable) definition of these quantities in the quantum domain~\cite{talknerAspectsQuantumWork2016,perarnau-llobetNoGoTheoremCharacterization2017,strasbergQuantumInformationThermodynamics2017,cerisolaUsingQuantumWork2017,niedenzuConceptsWorkAutonomous2019,beyerWorkExternalQuantum2020b,silvaQuantumMechanicalWork2021, beyerJointMeasurabilityNonequilibrium2022}.
Still, a vast amount of proposals for quantum heat engines and refrigerators have been investigated in recent years~\cite{levyQuantumAbsorptionRefrigerator2012,newmanPerformanceQuantumHeat2017,hayashiMeasurementbasedFormulationQuantum2017a,elouardExtractingWorkQuantum2017,mohammadyQuantumSzilardEngine2017,vonlindenfelsSpinHeatEngine2018,Wiedmann2020Mar,Wiedmann2021Jun, KoyanagiLawsThermodynamicsQuantum2022,KoyanagiNumericallyExactSimulations2022, kaneyasuQuantumOttoCycle2023}.
Importantly, all of those obey the Carnot bound and are thus constrained by the same limits we know from classical thermodynamics.
Examples of quantum heat engines which seem to break the Carnot limit rely on some sort of non-thermal energy reservoir~\cite{klaersSqueezedThermalReservoirs2017,thomasThermodynamicsNonMarkovianReservoirs2018,Binder2018}.

Even though quantum heat engines will most likely not be able to outperform classical ones, 
there are microscopic setups where quantum effects play an important role for the description of the thermodynamic behavior.
For example, it has been shown that coherences in the work medium or non-classical correlations can have a great influence on the performance of quantum engines, increasing power or efficiency in some scenarios~\cite{camatiCoherenceEffectsPerformance2019,cakmakErgotropyCoherencesOpen2020,mayoCollectiveEffectsQuantum2022,brunnerEntanglementEnhancesCooling2014a,francicaDaemonicErgotropyEnhanced2017,beyerSteeringHeatEngines2019,bresqueTwoQubitEngineFueled2021}, while resulting in a quantum disadvantage in others~\cite{brandnerUniversalCoherenceInducedPower2017,pekolaSupremacyIncoherentSudden2019}. 

As a complex many-body quantum system, a first principle theoretical treatment of a quantum heat engine based on a full system-bath model is notoriously difficult.
Regarding the reduced dynamics of the work medium, a variety of exact approaches has matured over the last years, steadily extending the accessible parameter regime~\cite{WangMultilayerMulticonfigurationTimeDependent2015, HartmannExactOpenQuantum2017, Strathearn2018Aug, TanimuraNumericallyExactApproach2020, XuTamingQuantumNoise2022, GeraSimulatingOpticalLinear2023, LinkOpenQuantumSystem2023}.
Nonetheless, to address the relevant properties such as efficiency and power of a quantum heat engine, a handle on the bath dynamics and its observables is needed in addition to the sole system dynamics.
In particular, the change of the bath energy and the energy attributed to the interaction Hamiltonian are of relevance. 

 Advanced methods for open system quantum dynamics allow for access to both quantities. Most notably, bath related observables can be obtained numerically from the auxiliary density operators of the \ac{HEOM} formalism of open system dynamics~\cite{ZhuExplicitSystembathCorrelation2012, TanimuraReducedHierarchicalEquations2014a, KoyanagiLawsThermodynamicsQuantum2022, KoyanagiNumericallyExactSimulations2022, Kato2016Dec}. Alternatively, a stochastic Liouville–von Neumann equation approach to open system dynamics~\cite{StockburgerMak1999} can be adapted for time dependent modulations and the determination of interaction and change of bath energy. It was used by Wiedmann {\it et al.}\ in \cite{Wiedmann2020Mar} to study the dynamics, power and efficiency of an Otto engine with an oscillator work medium, serving as a guidance for our treatment of the Otto cycle here, based on HOPS. It is also known that knowledge of appropriate system multi-time correlation functions allows for the determination of relevant bath properties~\cite{GribbenUsingEnvironmentUnderstand2022}.
In practice, this is numerically hard too, as it requires the evaluation of the full influence functional or process tensor~\cite{PollockNonMarkovianQuantumProcesses2018}, nowadays obtained from tensor network contraction methods~\cite{JorgensenExploitingCausalTensor2019, LinkOpenQuantumSystem2023}.

In this work, we show how to employ the stochastic \ac{HOPS} approach to open quantum system dynamics~\cite{Suess2014Oct, HartmannExactOpenQuantum2017, Hartmann2021Aug} for a full dynamical treatment of a quantum heat engine and its baths. 
Notably, HOPS reflects a stochastic representation of the full state of system and environment and thus, it comes as no surprise that collective bath observables are accessible, too. We show that they can be calculated in terms of the auxiliary states of the \ac{HOPS}, complementing similar findings in \ac{HEOM}~\cite{TanimuraReducedHierarchicalEquations2014a, Kato2016Dec, KoyanagiLawsThermodynamicsQuantum2022,latuneCyclicQuantumEngines2023}. Note that our HOPS formalism does not impose any constraints on the spectral density of the baths or the nature of the work medium, unlike the approach in Ref.\ \cite{Wiedmann2020Mar}, which is based on the infinite high energy cutoff limit. Here, we choose a bath frequency cutoff comparable to system energy scales.

Special features of the stochastic approach, e.g.\ the unitary stochastic representation of finite temperature, are shown to be of great benefit, as the hierarchy itself is determined from zero temperature properties only. Furthermore, all the advantages of HOPS are preserved. Foremost among them is the improved memory efficiency that results from working with pure state trajectories of dimension \(D\) of the system Hilbert space instead of density matrices of dimension \(D^{2}\).
Importantly, in this article we work out how to apply the non-linear variant of \ac{HOPS} to the study of system-bath interaction and bath observables, which allows to efficiently treat the regime of strong system-bath coupling.

As a central part of this work, we use HOPS to investigate the power and efficiency of a quantum heat engine from first principles, similar to the studies performed in Refs.~\cite{Wiedmann2020Mar,KoyanagiNumericallyExactSimulations2022}. 
A general model of a heat engine consists of a work medium \(\sys\) with Hamiltonian \(H_{\sys}\) which is coupled to \(N\) baths with bare Hamiltonians \(H_{\bath}^{(n)}\). The interaction is mediated by Hamiltonians \(H_{\inter}^{(n)}\).
Note that in general -- and in particular for engines that work in cycles -- both the system Hamiltonian and the system-bath interaction are explicitly time-dependent. Thus, the general global Hamiltonian that we consider in this study has the form
\begin{equation}
  \label{eq:openSystemHamiltonian}
  H(t) = H_{\sys}(t) + \sum_{n=1}^N \qty[H_{\inter}^{(n)}(t) + H_{\bath}^{(n)}],
\end{equation}
with bosonic heat baths.
On this global level, all energy contributions or their changes are well-defined and accessible by the \ac{HOPS}.
By taking a global unitary approach, difficulties often encountered in reduced descriptions, 
such as the ambiguity in the splitting of the interaction energy in a system and a bath contribution, do not occur.
We will rely on the well-defined expectation values of the respective parts of the Hamiltonian: \( \langle H_\sys \rangle \), \( \langle H_\inter \rangle \), and \( \partial_t \langle H_\bath \rangle \) (see also Refs.\cite{Wiedmann2020Mar, KoyanagiNumericallyExactSimulations2022}). No ad hoc separation of work-like and heat-like energy is called for.
\begin{figure}
\centering
\includegraphics[width=\columnwidth]{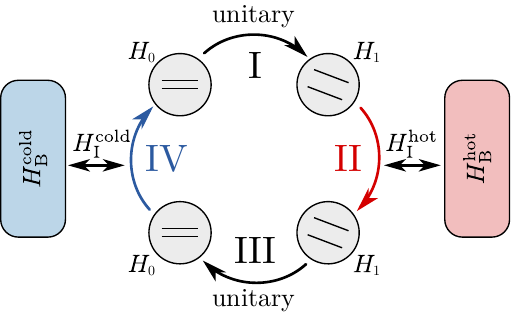}
\caption{The Otto cycle {exemplified with} a single two-level system as work medium. In general, the unitary ``compression'' (\phase{1})/``expansion'' (\phase{3}) of the work medium increases/decreases the level spacing while changing the orientation of the qubit, too.}
\label{fig:otto_cycle_scheme}
\end{figure}


We will illustrate the strength of such a global approach by studying a common model of a quantum heat engine using HOPS: the Otto cycle. A thorough and related analysis with an oscillator instead of our qubit work medium was performed by Wiedmann et  al.~\cite{Wiedmann2020Mar} based on a stochastic Liouville–von Neumann equation approach. 
Simplified investigations of the Otto engine in terms of master equations can be found in Refs.~\cite{Geva1992Feb,Karimi2016Nov,camatiCoherenceEffectsPerformance2019}. 
An Otto cycle consists of four separate \emph{phases}, which are depicted in Fig.~\ref{fig:otto_cycle_scheme} and listed in the following (consider the system initialized in equilibrium with the cold bath). 
\begin{description}
\item[\(\overset{\text{\rom{1}}}{t_0\to t_1}\) isentropic ``compression''] At time \(t=t_0\), the cold bath is decoupled from the system, that is, {$H^{\cold}_\inter(t_0)=0$}. Now, the system Hamiltonian changes from \(H_\sys(t_0) = H_0 \to H_\sys(t_1) = H_1\) and the work medium evolves unitarily. Depending on these two system Hamiltonians, the machine will work either as a heat engine or as a refrigerator. To operate in the heat-engine regime, the level-spacing of \(H_{1}\) should be greater than that of \(H_{0}\). This ensures that energy can be extracted from the system through inverting this step later in the cycle. 
The term ``compression'' is chosen as the level spacing \emph{increases}. This is analogous to the spectrum of a particle in a box whose volume is decreased.

\item[\(\overset{\text{\rom{2}}}{t_1\to t_2}\) isochoric heating]  The system Hamiltonian is kept constant. The interaction {$H_\inter^\hot$} with the hot bath is turned on and the system equilibrates with this reservoir.
  
    \item[\(\overset{\text{\rom{3}}}{t_2\to t_3}\) isentropic ``expansion''] The hot bath is decoupled,  {$H_\inter^\hot(t_2)=0$}, and the system Hamiltonian changes back to its initial value, i.e.\ \(H_\sys(t_2) = H_1 \to H_\sys(t_3) = H_0\).
    The energy of the entire system decreases. Ideally, the energy decrease is larger than the energy increase in phase \phase{1}.
    \item[\(\overset{\text{\rom{4}}}{t_3\to t_0}\) isochoric cooling] The system Hamiltonian is kept constant. The interaction with the cold bath {$H_\inter^\cold$} is turned on, the system equilibrates with the cold bath and, finally, the interaction is turned off. The cycle has closed and can be repeated.
\end{description}

The clear-cut separation of the four phases is an idealization.  In a
real Otto engine, be it classical or quantum-mechanical, all
operations are continuous in time. Moreover, the actions that are
performed in the four phases may overlap in time.  In fact, this
additional freedom will be exploited later to optimize the heat
engine.


Crucially, cyclic modulations will in general lead to non-equilibrium dynamics in the system.
Still, in (classical) macroscopic engines -- due to the many degrees of freedom and their fast internal thermalization -- the dynamics of the work medium is well described by quasistatic state changes.
Heat engines based on current quantum technologies, by contrast, use work media that are very small -- most notably qubits or single bosonic modes~\cite{anExperimentalTestQuantum2015a, zhangExperimentalDemonstrationWork2018, jalielExperimentalRealizationQuantum2019,Kurizki2021Dec,joshiExperimentalInvestigationQuantum2022, jiSpinQuantumHeat2022}.
For such microscopic systems, a quasistatic description is no longer appropriate. 
Foremost, the assumption of an instantaneous internal equilibration does not hold, in particular for finite work media as considered here.
Second, even if the cycle is slow enough such that the system can always be considered in a steady state with respect to the instantaneous external parameters, this steady state is generally not given by a Gibbs state with respect to the instantaneous local Hamiltonian. 
Such a description is valid in an ultra-weak coupling regime only~\cite{Breuer2002Jun, Hartmann2020Jan}. 
However, in a generic scenario with a finite (or even large) coupling strength between the system and the baths, for the steady state of the system the interaction Hamiltonian has to be taken into account~\cite{ThingnaGeneralizedGibbsState2012, CresserWeakUltrastrongCoupling2021,Wiedmann2020Mar}.

Thus, in order to fully describe a quantum heat engine it is necessary to consider transient effects, dynamics well beyond the quasistatic regime, as well as finite coupling strengths with the baths.
As the \ac{HOPS} method can satisfy all these demands, we are able to determine the dynamics of the engine for an arbitrary driving of the work medium through a time-dependent system Hamiltonian, as well as arbitrary (continuous) protocols for coupling/decoupling to/from the heat baths through the time-dependent interaction Hamiltonians. Moreover, there is no restriction on the dimension of the work medium nor on a specific form of the spectral density.

In this article, we go beyond previous considerations of the \ac{HOPS} approach to open system dynamics by exploiting that the change of the bath energy and the interaction energy can be obtained from within that formalism naturally.
We will provide illustrative examples with incomplete thermalization and different coupling strengths that confirm the impact of smoothly time-dependent couplings to the cold and hot baths, as shown in \cite{Wiedmann2020Mar}.
We stress again that the work contributions associated with modulation of the interaction Hamiltonians, i.e.\ switching the coupling to the baths on and off, significantly contribute to the overall work and energy balance.
It turns out that these contributions are sensitive to slight changes of the modulation protocol, as are power and efficiency (see also \cite{Wiedmann2020Mar}). Thus, a correct characterization of a quantum heat engine requires a careful assessment regarding all details of the cycle. 
Furthermore, we extend these considerations by exploring overlapping cycle phases with an eye on optimizing the performance of the engine.
Remarkably, we find that the coupling/decoupling of the baths can contribute positively to the power-output of the machine.

The remainder of this article is structured as follows.
In Sec.~\ref{sec:energies} we review natural definitions of power and efficiency of a heat engine in terms of the global energy balance of system, interaction, and bath.
Sec.~\ref{sec:qubit-engine} introduces a qubit (two-level) heat engine with an \ac{OLC} modulation protocol, on which all results of this work are based. We discuss the dynamics of the work medium and the flow of the various energetic contributions. We then turn to the parameter dependence of power and efficiency of the engine.
In Sec.~\ref{sec:infl-mod-protocol} we analyze modulations beyond the \ac{OLC} by allowing overlapping cycle phases and continue in Sec.~\ref{sec:infl-cycle-length} with an investigation on how cycle length and coupling strength affect the performance of the machine.
In Sec.~\ref{sec:method}, we present details of the employed \ac{HOPS} method, focusing on how relevant bath observables can be computed.
We close with conclusions and an outlook in Sec.~\ref{sec:conclusion}.

\section{Energy balance, power, and efficiency}
\label{sec:energies}

In classical thermodynamics, the performance of a heat engine is quantified in terms of the heat that is transferred from and to the system and the work extracted during a cycle.
For quantum heat engines, work and heat are not easily defined in such a setting. Fortunately, in our global approach we have access to the total energy balance of system, interaction and bath. Thus, we do not need to rely on ambiguous definitions of work and heat for the subsystems.
Instead, power and efficiency can be defined solely through the well-defined expectation values of the involved Hamiltonians \cite{Kato2016Dec,Wiedmann2020Mar,KoyanagiNumericallyExactSimulations2022} and are introduced below to make this work self-contained.

It has been demonstrated by Wiedmann et al.\cite{Wiedmann2020Mar}\  that for small quantum systems, the energy associated with the system-bath interaction Hamiltonian needs to be studied carefully. While in an ultra-weak coupling scenario one simply neglects this contribution to the total energy balance, for the finite-time cyclic heat engine studied here, it turns out to play an important role.

\subsection{Power}
The global dynamics of system and bath evolves unitarily, i.e.\ without any entropy production. Thus, a global energy change can always be considered work. Accordingly, the total power is the rate of change of the global energy,
\begin{equation}
  \label{eq:power}
  \begin{aligned}
        P(t) &= - {\frac{\dd }{\dd t}}\ev{H(t)} = - \ev{\partial_{t}H(t)}\\
          &= \underbrace{\ev{-\partial_{t} H_{\sys}(t)}}_{P_{\sys}(t)} + \sum_{n=1}^N\underbrace{\ev{-\partial_{t} H^{(n)}_{\inter}(t)}}_{P_{\inter}^{(n)}(t)}
  \end{aligned}
\end{equation}
where the last equality follows from the time independence of the bath Hamiltonians (see~\cref{eq:openSystemHamiltonian}). 
The total power in \cref{eq:power} consists of individual contributions associated with respective Hamiltonians, namely the system power \(P_{\sys}(t)\) and, likewise, power contributions  \(P^{(n)}_{\inter}(t)\) from the coupling/decoupling of the baths. Note that a usable power \emph{output} corresponds to a positive value for the power, as the global system loses energy.

For the performance of a heat engine, one is usually interested in the cycle-averaged long-time behavior. 
The average power \(\bar{P}\) and the work \(W\) performed over one such \emph{limit cycle} are defined through
\begin{equation}
        \label{eq:5}
        \bar{P}\cdot \Theta = W = \lim_{k\rightarrow\infty} \int_{k \Theta}^{(k + 1)\Theta}P(t)\,\dd t,
\end{equation}
where \(\Theta\) denotes the length of the cycle.

\subsection{Efficiency}

In conventional thermodynamics, the efficiency is defined as a benefit to cost ratio. In this case, the benefit is the work output \(W\). The cost is the energy consumed from the hot bath. 
Recall that due to the infinite number of bath degrees of freedom, the bath energy is infinite.
Nonetheless, the {\it change} of the bath energy \(\Delta\ev{H_\bath^\hot}\) is finite and is accessible by the global \ac{HOPS} method.
Note that this energy change is not necessarily only heat-like. In Ref.~\cite{Wiedmann2020Mar} it is shown that in the periodic steady state this quantity can be decomposed into terms that stem from the change in the system energy due to the interaction and changes due to the modulation of the interaction. Note however, that care has to be taken with such an interpretation as these contributions are not entirely independent through the coupled dynamics of the entire, global system. 

As we consider a general scenario with time-dependent and strong couplings, the modulation of the interaction Hamiltonian will also perform work on the baths~\cite{elouardExtendingLawsThermodynamics2023}. 
We could not care less: clearly, the efficiency of an engine is determined by the net energy loss of the hot bath over one cycle that is converted into work, accounting for all energetic processes in the bath, be it heat or work~\cite{latuneCyclicQuantumEngines2023}.
Consequently, the efficiency of the engine is
\begin{equation}
        \label{eq:4}
        \eta = \frac{W}{-\Delta\ev{H_\bath^\hot}} = \frac{\Delta\ev{H}}{\Delta\ev{H_\bath^\hot}},  
\end{equation}
where $\Delta$ denotes the change of a quantity over one limit cycle of length \(\Theta\). 
We note in passing that the limit cycle condition ensures $\Delta\ev{H_{\sys}}=0$ and $\Delta\ev{H_{\inter}}=0$, such that $\Delta\ev{H}=\Delta\ev{H_\bath^\hot} + \Delta\ev{H_\bath^\cold}$ holds.

It can be shown that \cref{eq:4} is bounded by the Carnot efficiency {\(\eta_{C} = 1 - {T_\cold}/{T_\hot}\)} under the assumption that a limit cycle has been reached. This is generally true whenever all system quantities are time-periodic as for our model (see~\cite{Kato2016Dec}, the argument given there can be easily extended to time dependent couplings).
For an \ac{OLC} the efficiency is also bounded by the efficiency of an ideal Otto cycle using a qubit as work medium ${\eta_{\mathrm{Otto}} = 1 - {\omega_\cold}/{\omega_\hot}}$, where $\omega_i$ is the energy gap of the qubit during coupling to the hot and cold bath, respectively -- a result arising from rather general considerations \cite{HenrichQuantumThermodynamicOtto2007}.

\section{Qubit heat engine model and its dynamics}

\label{sec:qubit-engine}

In this section, we let a single qubit take the role of the work medium for the heat engine, which is coupled to an infinite bosonic hot and cold bath along the cycle in a spin-boson like fashion.
The total Hamiltonian of the Otto engine thus reads
\begin{equation}
  \label{eq:SBM}
  \begin{aligned}
    H = \frac{\Omega}{2}& \pqty{\vec{s}(t) \cdot \vec{\sigma} + s_0(t)\id}\\
                     &+ \sum_i \Biggl(\frac{\sigma_x}{2} h_{i}(t) \pqty{B_{i}+B_{i}^\dagger} + H_\bath^{(i)}\Biggr),
  \end{aligned}
\end{equation}
where \(i=\hot,\, \cold\).
With bosonic annihilation and creation operators  $a,a^\dag$, the bath Hamiltonians are $H_\bath^{(i)}=\sum_{\lambda_i} \omega_{\lambda_i} a_{\lambda_i}^\dag a_{\lambda_i}$, and the coupling agents read $B_i=\sum_{\lambda_i} g_{\lambda_i} a_{\lambda_i}$. 
The explicit time dependence of both, system and system-bath interaction Hamiltonian are encoded in the real-valued smooth and periodic (period $\Theta$) functions $\vec s(t)$, $s_0(t)$ and $h_i(t)$.
Their choice determines the specific protocol for system modulation and bath coupling/decoupling.
The time-dependent identity contribution to the system Hamiltonian is introduced to control the ground state energy of the work medium (fixed to zero, in most cases). A similar scheme with, however, an instantaneous instead of a smooth switching on/off of the interaction Hamiltonian has been studied in Ref.~\cite{latuneCyclicQuantumEngines2023}.

In order to model a somewhat generic scenario for the qubit work medium, we choose
\begin{equation}
    \label{eq:system_modulation}
    \begin{gathered}
          s_0(t) = s_z(t) = 1 + f(t),  \\ s_x = \text{const.} < \frac{1}{2}, \quad \text{and} \quad s_y = 0,
    \end{gathered}
\end{equation}
where $0\le f(t) \le 1$ is a smooth function modulating the qubit energy gap periodically in line with the phases of the cycle (see also \cref{fig:otto_like_cycle}).
In terms of the classical Otto cycle, this system modulation corresponds to expansion and compression of the work medium.
In the expanded phase at $f(t)=0$ the (smaller) level spacing is $\varepsilon_0 = \Omega\sqrt{1 + (s_x/\Omega)^2}$ whereas in the compressed phase at $f(t) = 1$, we have (the larger) $\varepsilon_1 = 2\Omega\sqrt{1 + (s_x/2\Omega)^2}$.

Whenever $s_x \neq 0$, the time dependence of $f(t)$ results in a time-dependent orientation $\vec s(t)$ of the system Hamiltonian.  This implies that the instantaneous eigenstates of the system Hamiltonian become time-dependent -- the thermodynamic relevance of which will be discussed later. 
For $s_x=0$, by contrast, the eigenstates of $\sigma_z$ are the time-independent eigenstates of $H_{\sys}(t)$ throughout the cycle.

With $B_i(t)$ denoting the above coupling agent in the interaction picture with respecfunctiont to its $H_\bath^{(i)}$, the microscopic structure of the two baths and their coupling to the work medium is captured in the zero-temperature bosonic bath correlation function \cite{Weiss2012}

\begin{equation}\label{bcf}
  \begin{aligned}
      \alpha_i(t-s) &{}\equiv
                 \langle 0 | B_i(t) B_i^\dag(s) | 0 \rangle = \sum_{\lambda_i} |g_{\lambda_i}| \eu^{-\mathrm{i}\omega_{\lambda_i}(t-s)} \\
               & = \frac{1}{\pi}\int_0^\infty d\omega J_i(\omega) \eu^{-i\omega(t-s)}.
  \end{aligned}
\end{equation}

We choose an Ohmic spectral density $J_i(\omega)$ for each bath with identical exponential cutoff (see Methods Sec.~\ref{sec:method} for details) but possibly different coupling strengths $\tilde \eta_i$,
\begin{equation}
        \label{eq:ohmic_sd}
        J_i(\omega) = \tilde \eta_i\, \omega \eu^{-\frac{\omega}{\omega_{c}}}.
\end{equation}
The cutoff frequency $\omega_c$ is set to the rather small value $\omega_c = \Omega$ of the order of the qubit level spacing. Thus, the memory time of the environment is of the same order as the system timescale, and the dynamics is expected to be non-Markovian.
Unless otherwise stated, we set the temperature of the cold bath to $T_\cold = \Omega/2$ and that of the hot bath to $T_\hot = 4\Omega$.
The latter is such that a reasonable occupation of \(p_{\uparrow} \approx 0.38\) of the upper level can be expected from the corresponding Gibbs state of the qubit (see also lower panels of Fig.~\ref{fig:otto_like_cycle}).
The lower temperature (up state occupation of \(p_{\uparrow} \approx 0.12\)) is small enough to yield a significant system energy difference during one cycle.
Note that we refer to the Gibbs state here as a rough estimate for the choice of temperatures --
the true quasi-equilibrium state while coupled to a bath during the cycle will in general not be the local Gibbs state due to the strong interaction.

In order to achieve a comparable effect of both baths on the system dynamics, we choose the coupling strengths $\tilde \eta_i$ such that the value of the effective thermal spectral densities agree at the approximate level spacing, i.e.\ $J_{\beta_\cold}(\Omega) \stackrel{!}{=} J_{\beta_\hot}(2\Omega)$ with
\begin{equation}
  \begin{gathered}
    J_{\beta_i}(\omega) = \frac{J_i(\omega)}{1 - \eu^{\beta_i \omega}},\\
    J_i(\omega) := -J_i(-\omega) \; \text{for} \; \omega < 0 \;,
  \end{gathered}
\end{equation}
and inverse temperature $\beta = (kT)^{-1}$.
This choice is motivated from the ultra-weak coupling regime where, according to Fermi's Golden Rule, the environmental influence is mainly determined by the value of the thermal spectral density at the resonance frequencies~\cite{Breuer2002Jun}.
The specific value
\begin{equation}
    \delta := \frac{J_{\beta_\hot}(2\Omega)}{\Omega} \equiv \frac{J_{\beta_\cold}(\Omega)}{\Omega} = \tilde \eta_\cold \frac{ \eu^{-\frac{ \Omega}{\omega_c}}}{1 - \eu^{\beta_\cold \Omega} }
    \label{eq:cpl_delta}
\end{equation}
serves as global measure for the coupling strength between the qubit and the baths.
Unless otherwise stated, we choose \(\delta =0.7\).

For the following calculations, the initial condition is fixed to the product state
\begin{equation}
  \rho_0 = \ket{\downarrow}\bra{\downarrow} \otimes \rho_{\beta_\cold} \otimes \rho_{\beta_\hot},
\end{equation}
where $\ket{\downarrow}$ denotes the eigenstate of $\sigma_z$ with eigenvalue $-1$ and $\rho_\beta$  the Gibbs state of the corresponding bath Hamiltonian. As the system starts off in a pure state, we expect some transient dynamics before the heat engine reaches a periodic limit cycle (the third, for our choice of parameters).

\subsection{Otto-like cycle modulation protocol for the qubit engine}

\begin{figure*}
    \centering
    \includegraphics[width=\textwidth]{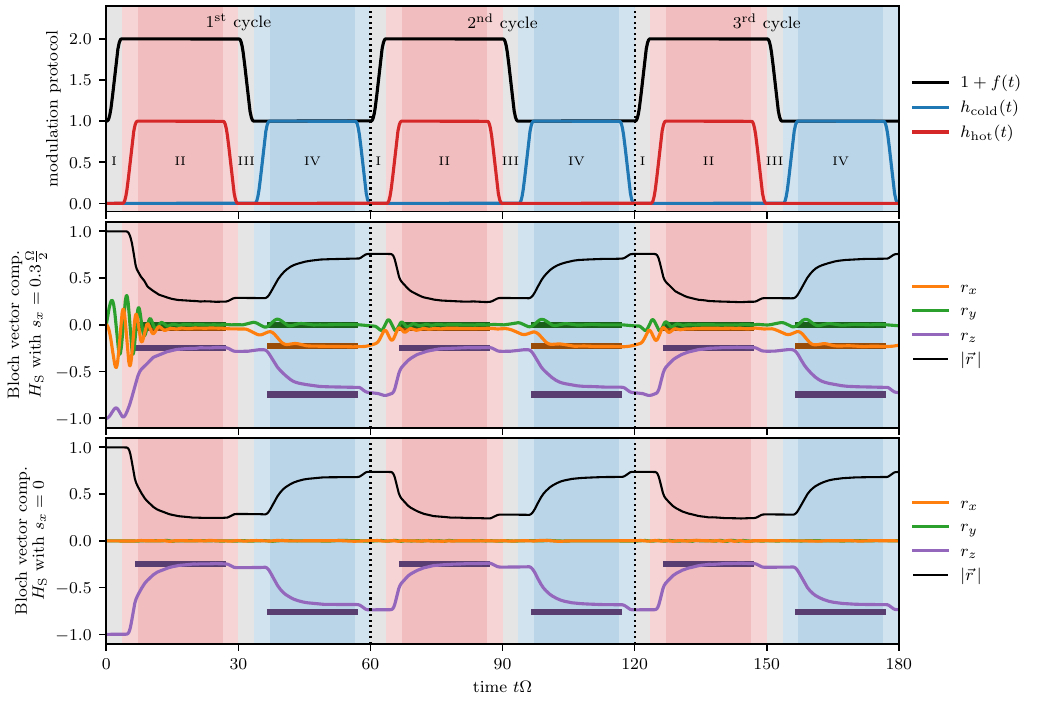}
    \caption{The modulation protocol\cite{Wiedmann2020Mar} (upper panel) of the Otto-like cycle (\ac{OLC}) shows the separated cold (blue line and background) and hot (red line and background) phases. The system is modulated (black line) during the phases where the interaction with both baths is off.
    For that protocol, the reduced dynamics of the qubit in terms of the Bloch vector is shown in the middle ($s_x = 0.3\Omega/2$) and the lower panel ($s_x = 0$).
    The length of the Bloch vector (indicating the purity of the state) is shown in black.
    The horizontal thick lines correspond to the Bloch vector components of the Gibbs state of the qubit with respect to the current system Hamiltonian and the temperature of the corresponding bath.
    The vanishing $x$ and $y$ components of the Gibbs state in the lower panel are not shown
    (parameters: $\delta = 0.7$, $\omega_c=\Omega$, $T_\mathrm{cold}=\Omega/2$, $T_\mathrm{hot}=4\Omega$, $\Theta=60/\Omega$, $\tau_\mathrm{tr} = 0.06\Theta$).
    }
    \label{fig:otto_like_cycle}
\end{figure*}

We begin our investigation with a modulation protocol inspired by Ref.~\cite{Wiedmann2020Mar} which aims to be close to the ideal Otto cycle (see upper row in Fig.~\ref{fig:otto_like_cycle}), i.e.:
\begin{description}
    \item[\phase{1} Compression phase] Change the system Hamiltonian from expanded $f=0$ to compressed $f=1$ (see Eq.~\eqref{eq:SBM} and Eq.~\eqref{eq:system_modulation}), while the interaction with the baths is off ($h_i(t)=0$).
    \item[\phase{2} Hot Phase] Turn on, hold and turn off the interaction with the hot bath using a smooth, non-zero $0\leq h_{\hot}(t)\leq 1$.
    \item[\phase{3} Expansion phase] Undo the change of the system Hamiltonian (suitable $f(t)$).
    \item[\phase{4} Cold phase] Turn on, hold and turn off the interaction with the cold bath (through $0\leq h_{\cold}(t)\leq 1$). Restart the cycle.
\end{description}
We call such a protocol \acf{OLC} to emphasize that the four phases are non-overlapping, as in the Otto cycle, while any modulation of the global Hamiltonian is modelled by a smooth transition.
In particular, we utilize smooth sigmoid-like functions that are twice continuously differentiable.
The time interval to perform a single transition will be denoted by $\tau_\mathrm{tr}$ and is initially set to $\tau_\mathrm{tr} = 0.06\Theta$.
The cycle time $\Theta$ is chosen to be \(\Theta = 60/\Omega\), which allows the system to relax to an effective steady state during the phase where it is in contact with either of the baths (given that the coupling strength is no less than roughly $\delta \sim 0.5$, see Eq. \eqref{eq:cpl_delta}).  This also means that we consider a regime where the correlation time of the bath ($\sim 1/\Omega$, see above) is short compared to the cycle length. These parameters are chosen to be in a regime of positive power output without too much fine-tuning.

\bigskip
\begin{figure*}
  \centering
  \includegraphics[width=\textwidth]{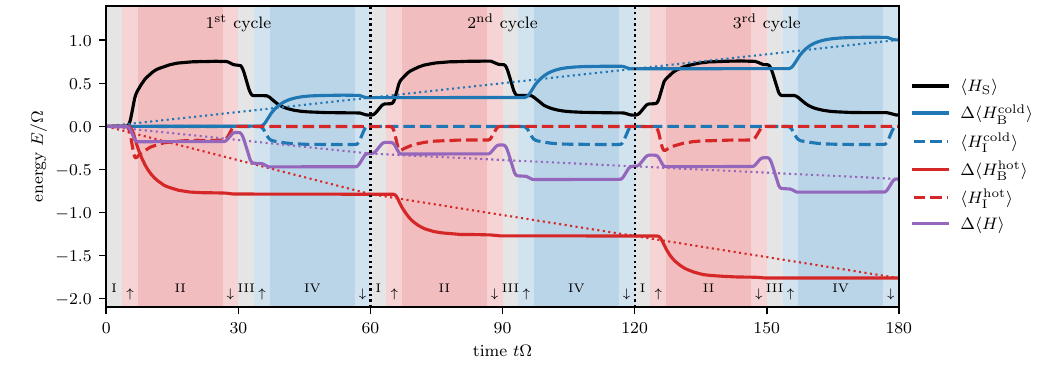}
  \caption{
    Dynamics of the various energy contributions:
    work medium (system) energy (black),
    change of the cold bath energy (blue),
    cold bath interaction energy (dashed blue),
    change of the hot bath energy (red),
    hot bath interaction energy (dashed red) and
    change of the global energy (purple - not constant due to the explicit time dependence of the global Hamiltonian).
    The colored dotted lines show the overall trend for the respective quantity after a full cycle
    (same parameters as in Fig.~\ref{fig:otto_like_cycle} but with $s_x=0$ only).
  }
  \label{fig:energy}
\end{figure*}

Before focusing on power and efficiency, we first discuss the dynamics of the work medium for such an \ac{OLC}, visualized in Fig.~\ref{fig:otto_like_cycle}.
In particular, the evolution of the Bloch vector
\begin{equation}
    \vec r = (r_x,r_y,r_z) = \mbox{Tr}[\vec\sigma \rho_{\sys}]
\end{equation}
is shown for two different choices of the modulation of the system Hamiltonian in the middle and lower panel of that figure.

First, the middle panel shows the more general case where the scale \emph{and} the orientation (\(s_{x}\neq 0\)) of the qubit Hamiltonian is modulated.
During the initial compression phase \phase{1} (grey background), the qubit state rotates on the surface of the Bloch sphere due to the $\sigma_x$ component of the system Hamiltonian. This rotation is visible as initial coherent oscillations of the $x$ and $y$ Bloch vector components.
At the beginning of phase \phase{2}, the interaction with the hot bath is smoothly turned on (light red background).  
As expected \cite{LeggettDynamicsDissipativeTwostate1987}, the qubit loses purity, as is indicated by the decrease of the absolute value of the Bloch vector (black solid line, middle panel of Fig.~\ref{fig:otto_like_cycle}).
Once the interaction is fully turned on (red background) the qubit relaxes towards an effective steady state.
In case of the hot bath, this steady state agrees fairly well with the local Gibbs state $\sim \eu^{-\beta_\hot H_S(f)}|_{f=1}$ of the qubit at the hot bath temperature (thick red horizontal lines).
Notably, during the decoupling from the hot bath (light red background at the end of phase \phase{2}) the qubit experiences a slight change towards a purer state.
This is closely related to a small decrease of the system energy during the decoupling stage (further discussed in the next paragraph).
The isolated system modulation in phase \phase{3} (grey background) produces unitary dynamics (constant length of the Bloch vector).
Since the orientation of the system Hamiltonian changes, its instantaneous eigenstates change as well, as does the qubit state.
As soon as the interaction with the cold bath becomes active (blue background) the qubit state becomes purer which indicates its cooling.  The qubit then reaches an effective steady state while in contact with the cold bath. Yet, that state deviates markedly from the corresponding local Gibbs state (thick horizontal blue lines). As our choice of parameters leads beyond the usual weak coupling regime, this is to be expected and shows the influence of the interaction Hamiltonian on the steady state.
Again, turning off the interaction with the bath results in a slight increase of purity.
Now the cycle is repeated, and we see that the initial transient dynamics -- which depended on the chosen initial qubit state -- are already strongly suppressed at the start of the second cycle.
There is no noticeable difference between the second and third cycle. Consequently, we can regard the third cycle as a reasonable approximation of the limit cycle.

Now we consider the second, somewhat simpler scenario, where we set $s_x = 0$ in \cref{eq:system_modulation}. Here, the modulation keeps the system Hamiltonian diagonal in the $\sigma_z$-basis for all times.
In that case, it holds generally true that the spin-boson Hamiltonian acts on two distinct subspaces of the global Hilbert space and, thus, the Schrödinger equation does not induce any transitions between them (parity symmetry).
As a consequence, our choice of initial condition results in dynamics confined to the $z$-axis of the Bloch sphere.
This behavior can be seen in the computed simulations shown in the bottom panel of Fig.~\ref{fig:otto_like_cycle}.
We note, however, that the individual pure state trajectories of the employed stochastic \ac{HOPS} method (details in Sec. \ref{sec:method}) do not share that property.
This is why very small deviations of $r_x$ and $r_y$ from zero show up in the simulations. These deviations can in principle be made arbitrarily small by increasing the number of stochastic trajectories used to calculate the reduced state.
We thus conclude that in the lower panel for $s_x=0$, the initial transient dynamics is very different from the middle, and moreover, $r_x$ and $r_y$ remain zero for all times.
Still, the overall limit cycle behavior, featuring the relaxation towards an effective steady state with higher energy when in contact with the hot bath and lower energy when in contact with the cold bath, is very similar in both cases.
Therefore, in the following discussion on power and efficiency, we will focus on that simpler case without the static $\sigma_x$ contribution.

\bigskip
So far we have discussed the exact dynamics of the work medium, i.e.\ of the reduced qubit state. Considering the explicit time dependence, the strong coupling to the environment, and not least a bath correlation time which is of the order of the system timescale, this is challenging on its own.
Additionally, the \ac{HOPS} approach allows access to the dynamics of the bath degrees of freedom, to which we turn next. In particular, changes of bath energies as well as the energetic contribution due to the interaction Hamiltonian (details in App.~\ref{sec:bath-energy-change} and App.~\ref{sec:interaction_energy}) can be computed. Thus, we are in a position to calculate power and efficiency of the quantum heat engine from first principles.

First, we show in Fig.~\ref{fig:energy} the dynamics of the individual energy contributions while operating the engine. Using HOPS, the detailed access to the dynamics of all energy contributions, including the interaction energy, allows us to present results beyond the earlier treatment in Ref.\ \cite{Wiedmann2020Mar}. 
The system energy $\langle H_\sys \rangle$ (black) and both interaction energies $\langle H_\inter^\mathrm{c./h.} \rangle$ (dashed blue, dashed red) are shown in terms of their absolute value.
For the infinite baths, it only makes sense to consider energy
changes. We choose $\Delta \langle H_\bath^\mathrm{c./h.} \rangle$ (solid blue, solid
red) to denote the change of the bath energy from the value at their
respective initial conditions.
\begin{description}
\item[{Initial system compression} (\phase{1})] Since the qubit is initially prepared in the local ground state with zero energy, all quantities are zero.
  The pure compression during the first system modulation (gray shading) only lifts the upper qubit level, which leaves the system energy (solid black) at zero (referring also to the dynamics shown in the bottom panel of Fig.~\ref{fig:otto_like_cycle}).
  
\item[{Hot phase} (\phase{2})]
When turning on the interaction with the hot bath (\phase{2}$\uparrow$, light red shading), the energy of the system rises while the contributions of the interaction and the bath decrease.
Summing up the three parts shows that the change in total energy $\Delta \langle H \rangle$ (purple) decreases, too.
This means that while switching on the coupling to the hot bath, work is extracted (in contrast to the standard thermodynamic Otto-cycle).
Once the interaction with the hot bath is fully established (red shading) the total energy remains constant as the global Hamiltonian is now time-independent.
Powered by the energy flow from the hot bath, the system heats up further as the qubit relaxes towards its temporary (hot) steady state. Notice,
however, that part of that missing hot bath energy flows into the interaction energy, which increases slightly after the initial sharp drop.
Before the interaction with the hot bath is switched off, the dynamics of the energy contributions have relaxed and reach constant values.
The process of switching off the hot bath (\phase{2}$\downarrow$, light red shading) requires energy since the interaction energy goes from its temporarily negative value to zero.
That energy is only partly provided by the system (and, to a lesser degree, by the hot bath), meaning that the total energy rises: switching off the hot bath requires work.

\item[{System expansion} (\phase{3})]
The decrease of the system energy upon expansion of the hot qubit directly carries over to a decrease of total energy, i.e.\ further work is extracted. This is clearly the main mechanism for work extraction (as for the standard Otto-cycle that motivated the design of the model).

\item[{Cold phase} (\phase{4})]
Switching on the interaction lowers the total energy and some work is extracted (\phase{4}$\uparrow$, light blue shading).
This means that in absolute values, the energy decrease of the system and the interaction is larger than the increase of the cold bath energy.
When fully coupled (blue shading) the cold bath takes energy from the system. The qubit is cooled as intended. At the same time -- but to a lesser degree -- the cold bath also takes up energy from the interaction.
Again, the process of decoupling the bath requires work (\phase{4}$\downarrow$, light blue shading).
The increase of the interaction energy is mainly powered by that external work, but receives small contributions from the system and the cold bath, too.
Overall, the full cold phase requires work.

\item[{\(2^{\text{nd}}\) system compression} (\phase{1}.2)]
In contrast to the initial compression, as the qubit has not completely relaxed to its ground state, the compression of the system now requires some work, too.
After that, the general behavior of the following cycles agrees with the first one.
\end{description}

Following the overall change of the bath energies over consecutive cycles (dashed lines in Fig.~\ref{fig:energy}) reveals that the energy flow into the cold bath is far less than the energy lost from the hot bath.
Therefore, part of that energy lowers the total energy which has been extracted as work due to the various time dependent modulations of system and interactions.
In other words, we have demonstrated that the \ac{OLC} based on the spin-boson model with Ohmic environments works as a heat engine. 
Note that once the system has reached its limit cycle, the change of system energy over one cycle is zero.
By construction, this holds true for the interaction energy, too.

In numbers, during the third cycle, the energy change is $\Delta \langle H_\bath^\mathrm{hot} \rangle  = -0.489 \Omega$ for the hot bath and $\Delta \langle H \rangle = -0.147 \Omega$ for the total energy.
This amounts to an average cycle power of $\bar P = 2.45 \cdot 10^{-3} \Omega^2$ and an efficiency of $\eta = 30.0 \%$. The first protocol in \cref{fig:otto_like_cycle}, with a $\sigma_x$-component of $s_x=0.3 \Omega/2$ of the system Hamiltonian, achieves \(\approx 93\%\) of the above mentioned power output and efficiency. Increasing \(s_{x}\) to \(0.5\Omega/2\) (not shown in \cref{fig:otto_like_cycle}) decreases the work output and efficiency further to \(\approx 80\%\) of the model with \(s_x=0\). The primary work extraction mechanism in the present model is the modulation of the Hamiltonian of the work-medium. If the modulated part of the system Hamiltonian does not commute with the constant part, coherences are generated. Thus, the Bloch vector of the work medium moves away from the axis along which energy is measured, leading to decreased work output for a similar change in hot bath energy. Therefore, the efficiency and power both change by the same relative amount. The decrease in power output due to coherences is known as ``quantum friction''~\cite{Binder2018,Feldmann2003}. 

Next, we address the sensitivity of these operationally relevant quantities with respect to the modulation protocol.

\section{Influence of the modulation protocol, overlapping cycle phases}
\label{sec:infl-mod-protocol}
\begin{figure}
    \centering
    \includegraphics[width=\columnwidth]{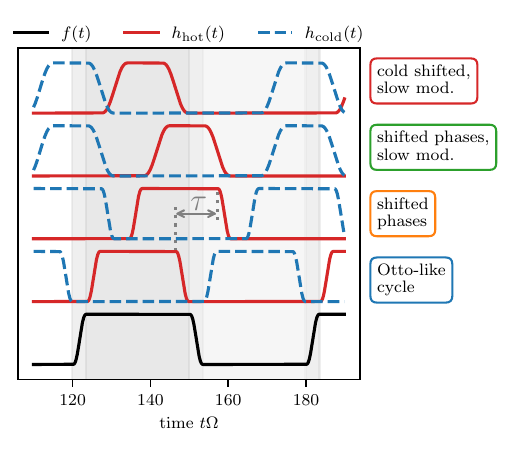}
    \caption{The different modulation schemes discussed in the
      text. In contrast to Ref.~\cite{Wiedmann2020Mar}, the bath
      modulation phases are shifted relative to the system modulation, and relative to
      each other. From the bottom up, the modulation of the
      \ac{OLC} is shown. This cycle features no shifts and all
      transitions happen within the time
      \(\tau_{\mathrm{tr}}=0.06\Theta\). Above, the bath-coupling modulation
      is shifted by the amount \(\tau\). Next, the modulation shift is
      the same as below, but with a slower transition time
      \(\tau_{\mathrm{tr}}=0.12\Theta\). In the top row only the cold bath
      coupling modulation is shifted (slowly as below).  The shading
      of the background corresponds to the modulation of the system
      shown in the lowest row. }
    \label{fig:fig_cycles}
\end{figure}
\begin{figure*}
        \centering
        \includegraphics[width=\textwidth]{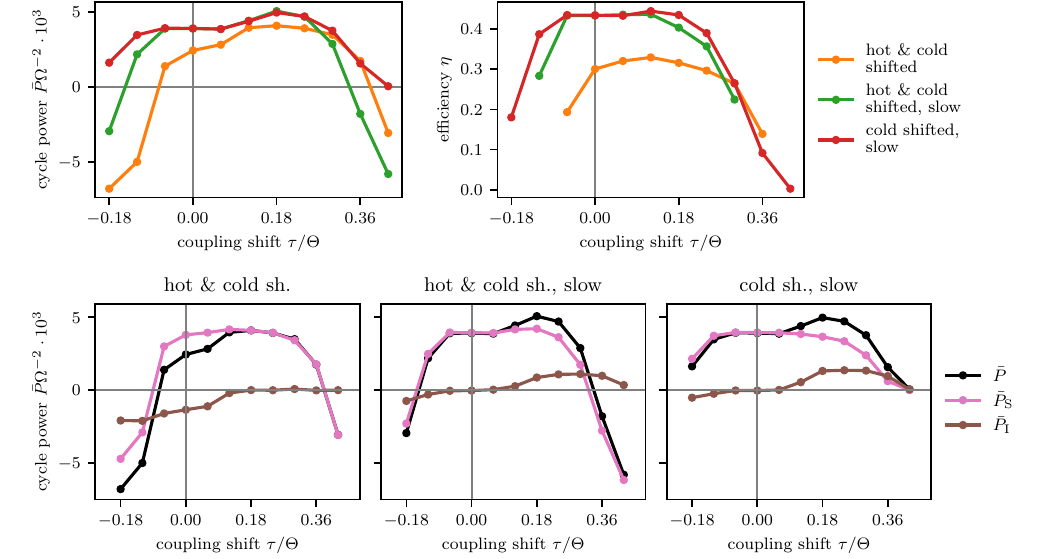}
        \caption{
        Upper row: Power and efficiency as a function of the shift of the timing of the coupling.
        A shift of zero (gray vertical lines) corresponds to the original \ac{OLC} protocol.
        We consider three cases: Hot and cold phases are both shifted (orange lines), they are both shifted yet the switching on to maximum coupling is slower (green lines), and the cold bath only is shifted with slower transition to maximum coupling (red lines).
        Efficiency is shown for scenarios with positive work only. 
        Bottom row: The individual contributions to the total average power (black lines) due to change in the system (magenta lines) and in the interaction Hamiltonian (brown lines) are shown. 
        The three diagrams correspond to the different cases of the upper row
        (parameters: $s_x=0$, $\delta = 0.7$, $\omega_c=\Omega$, $T_\mathrm{cold}=\Omega/2$, $T_\mathrm{hot}=4\Omega$, $\Theta=60/\Omega$, $\tau_\mathrm{tr} = 0.06\Theta$, slow modulation $\tau_\mathrm{tr} = 0.12\Theta$).
        }
        \label{fig:power_efficiency}
\end{figure*}
It is clear that the timing and modulation of the bath interactions will influence power and efficiency.
For instance, a refrigerator is realized by exchanging the hot and cold bath in the sequence of the cycle. 
A shift in time of the bath couplings may result in a temporal overlap with the expansion/compression phase: it turns out that the efficiency of the engine can be significantly enhanced by carefully tailoring such an overlap.
In the following we investigate the influence of the modulation
protocol by considering three cases: (i) a simultaneous shift in time
of the cold and hot bath interaction, (ii) that same simultaneous
shift but with slower transitions, and (iii) the case where only the
cold bath is shifted, with those slower transitions (see
\cref{fig:fig_cycles}). This third case allows for an overlap of the
hot and cold phases, with even further room for optimization. We also tried to modulate the various parts of the Hamiltonian simultaneously, yet didn't see further improvement of the performance.

\begin{figure*}
        \centering
        \includegraphics[width=\textwidth]{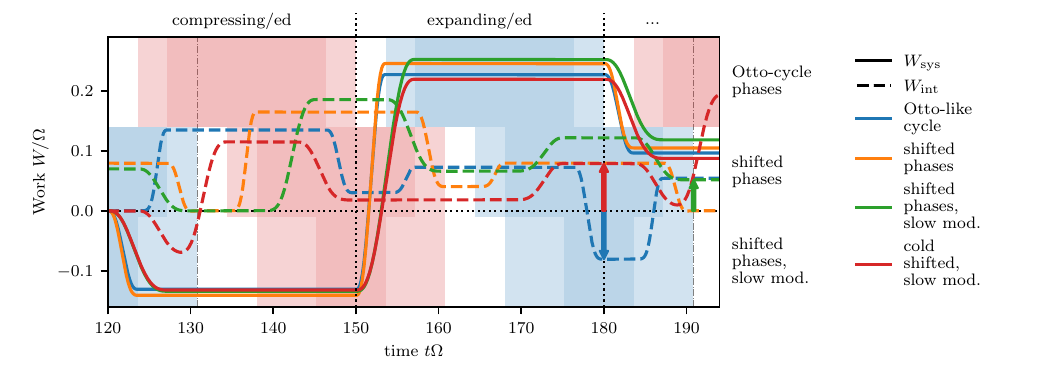}
        \caption{
        Work performed by the system (solid lines) and by the interaction modulation (dashed lines), shown as a function of time during a limit cycle and slightly beyond (same parameters as in Fig.~\ref{fig:power_efficiency}) for the best-performing cycles (maximum power/efficiency) with \(\tau=0.18\Theta\).
        As reference, we show again the work for the original \ac{OLC} modulation (upper shading of the phases) with blue lines.
        The blue arrow indicates that work needs to be done in order to drive the interaction with the baths.
        For the shifted modulation (orange lines) the compression/expansion of the qubit takes place when the system is in full contact with the cold/hot bath (see middle shading of the phases).
        In this case, the difference of interaction work after a full cycle is almost zero.
        For the shifted cycle and a slower modulation (green lines, lower shading of the phases) one actually gains work from the driving of the interaction, as emphasized by the green arrow. 
        When only shifting the cold phase (red lines, no shades for that protocol shown but see \cref{fig:fig_cycles}) the work gain due to the bath interactions is even larger, although the system work is smaller.
        Note that we compare the work curves of the cycles where both modulations are shifted just after decoupling from the cold bath (dash-dotted vertical lines).
        }
        \label{fig:work_shifted}
\end{figure*}

\begin{figure*}
        \centering
        \includegraphics[width=\textwidth]{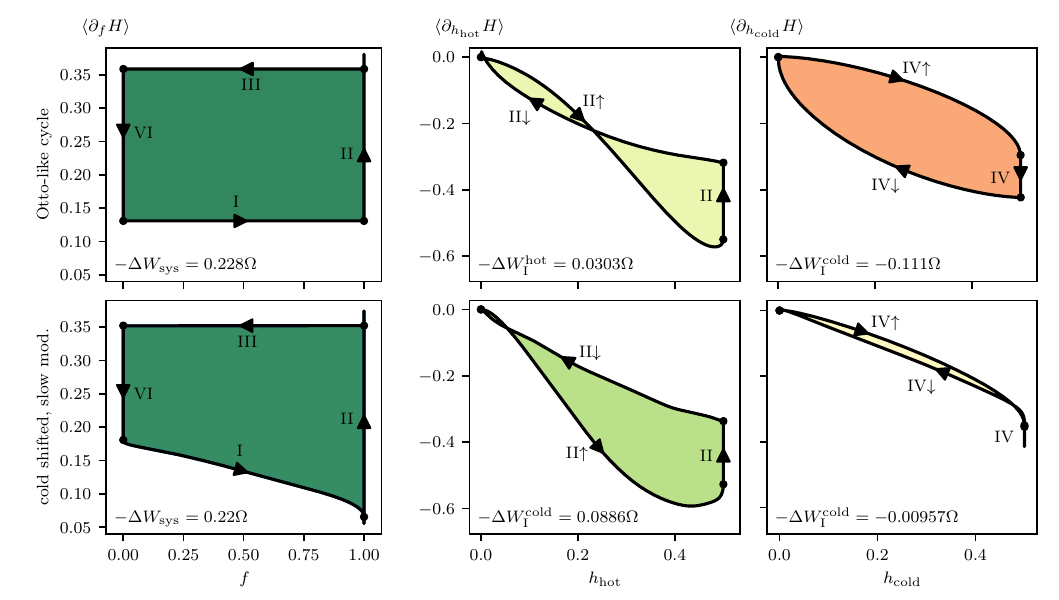}
        \caption{
          In correspondence with standard thermodynamic pressure-volume diagrams, the conjugate variables $\langle \partial_X H \rangle$ vs.\ $X$ are plotted for the limit cycle (see also \cref{eq:work_contributions}).
          Here, $X$ is either the system modulation $f$ (left column), or the interaction modulation $h$ with the hot (middle) and cold bath (right).
          From such a representation (negative) work amounts to the enclosed area. The coloring of the area is chosen according to the scale presented in \cref{fig:power_scan_linear}. Green shades correspond to positive (= usable) power output, whereas red shades indicate negative power.
          The upper row pertains to the \ac{OLC} whereas the lower row displays the work diagrams for the best performing slow cycle with \(\tau_{\mathrm{tr}}=0.12\Theta\) and with the cold bath modulation shifted by \(\tau=0.18 \Theta\) (see the rightmost lower panel in \cref{fig:power_efficiency}).
          The plots show that the slower modulation and the time-shift in the cold modulation reduces the energy cost of the hot phase and boosts the energy output from decoupling the hot bath.
          Some diagrams feature ``spikes'' (upper left, lower right).
          These appear when the expectation value $\langle \partial_X H \rangle$ overshoots the final value at the end of a phase. This is a token of the departure from standard thermodynamics. Take the left-most column as an example. When the hot bath is decoupled at the end of the hot phase (\phase{2}), the system energy is lowered slightly. This does not usually happen in standard thermodynamics, where the thermalization process is complete and monotonic.
          In addition, the ``kink'' seen in the lower middle panel is due to the onset of the cold phase before the hot phase has finished.
        \label{fig:pv}
      }
\end{figure*}

A common temporal shift of both couplings amounts to the replacement
$h_\mathrm{h./c.} (t) \rightarrow h_\mathrm{h./c.} (t - \tau)$ with the shift
parameter $\tau$.
In Fig.~\ref{fig:power_efficiency}, upper row, the power and efficiency of the heat engine as a function of the coupling shift $\tau$ relative to the modulation of the work medium is shown.
It turns out that a positive shift -- a delay of the hot/cold phases -- enhances the power up to a noticeable factor of $\approx 1.7$ at a shift of $\tau \approx 0.2 \Theta$ (orange line).
For larger shifts, the power drops and eventually becomes negative, which signifies the regime of a refrigerator.
As shown in the bottom row of Fig.~\ref{fig:power_efficiency}, left, the primary source for the boost in power is indeed the contribution of the interaction energy. This underlines the importance of considering that contribution to the power output in agreement with Ref.~\cite{Wiedmann2020Mar}, where under their choice of modulations, switching on the interaction always requires work.
Indeed, for the original \ac{OLC} protocol from above (lower left panel at $\tau=0$) we observe the same behavior (see Fig.~\ref{fig:work_shifted} for details). Now, however, we see that
this work can be reduced to zero, or work can even be extracted for a newly designed protocol with shifted modulation (overlapping phases).  Once the shift is too large, the compressed (expanded) state is no longer heated (cooled) and the work extraction due to the system modulation quickly drops.  This leads to a dramatic decrease in average power output and efficiency, as holds true for negative shifts, too.  Although the power shows a plateau over roughly $1/6$ of the cycle length, the efficiency starts to decline earlier, for shifts $\tau \gtrsim 0.2 \Theta$, as displayed in Fig.~\ref{fig:power_efficiency}, upper row, right. Note, however, that power \emph{and} efficiency can be improved by the shift in the modulation timing at the same time.

In addition to the timing of the modulation, we show in \cref{fig:power_efficiency} (green) that the speed of the modulation has a significant impact, too, as was found in Ref.\ \cite{Wiedmann2020Mar}.
As an example, we consider a protocol where all modulation steps are performed at half speed
({\it shifted, slow}), while keeping the cycle length constant.
\cref{fig:fig_cycles} as well as the background shades at the bottom of \cref{fig:work_shifted} depict the phases of that protocol.
The parameter $\tau_\mathrm{tr}$ is used to specify the duration of a single transition (``normal speed'': $\tau_\mathrm{tr} = 0.06\Theta$, half speed: $\tau_\mathrm{tr} = 0.12\Theta$ referred to as slow modulation).
The lower middle panel of Fig.~\ref{fig:power_efficiency} shows that the system contribution to the average power is nearly unaffected, while the interaction contribution is raised compared to the case with original speed.
Notably, for slow modulation, the average power contribution from the interaction becomes positive for a wide range of positive shifts with a maximum at $\tau \sim 0.25 \Theta$.
At $\tau = 0.18 \Theta$, the total power has increased by more than a factor of 2 compared to the original \ac{OLC}.
At the same time, slowing down the modulation results in a larger efficiency for the heat engine of more than $40\%$ compared to the value of $\sim 30\%$ for the original speed (Fig.~\ref{fig:power_efficiency}, upper row, right). For a detailed analysis of these findings, see Fig.~\ref{fig:work_shifted}, where the evolution of system and interaction work contributions for the different modulation protocols are displayed.

As an intuitive explanation for the observed improvement in the interaction power contribution, consider the following: In the expanded phase, the spectral density of the baths (recall that we have chosen the sames spectral density for both baths) has its maximum at the system level-spacing (\(\varepsilon_{0}= \Omega\)). This resonance leads to a larger magnitude of the \emph{negative} interaction energy. In the sudden limit, where the interaction is turned-off instantaneously, this energy would be entirely compensated by the interaction work. In the adiabatic limit, the opposite would be true, and the negative interaction energy would be entirely supplied by changes in system and bath energy. Our regime lies between those two extremes, and thus it requires work to decouple the bath. This work input grows with the magnitude of the interaction energy.
If this energy is reduced by any means other than external modulation, \emph{less} external work is needed to turn off the interaction. This is accomplished by detuning the system from the bath (compression) before decoupling.  The effect is also present for the hot bath. However, the effect should be \emph{reversed} compared to the cold bath. Recall, in the \emph{OLC}, the system is decoupled from the hot bath when it is compressed (\(\varepsilon_{1}=2\Omega\)) and thus off-resonant with the (zero-temperature) spectral density. Shifting the decoupling of the hot bath into the expanded phase actually brings the system in resonance. This doesn't seem to damage the performance of the cycle, which might be due to the higher temperature of the hot bath which obscures the zero-temperature structure of the bath through thermal fluctuations (see \cref{sec:non-zero_temperature}).

According to the above, not shifting the coupling to the hot bath should improve the performance of the engine even further.  Thus, we consider a set of protocols where the modulation of the interaction with the cold bath is shifted only (in combination with slow modulations). The results shown in Fig.~\ref{fig:power_efficiency} and Fig.~\ref{fig:work_shifted} (red lines) indicate that the mean interaction power is even larger, compared to a simultaneous shifting, but the system power decreases slightly.  In total, a very similar behavior of the average power is observed for values of the shift that are not too large.  However, for the protocol where the cold bath timing is changed only, the efficiency for the shift with the largest average power is increased from $40.3\%$ to $43.4\%$.

A more detailed analysis shows that the improvement in efficiency originates from the reduced energy flow out of the hot bath. If the hot bath is decoupled after the work medium is expanded, additional energy can flow into the system from the hot bath during and after the expansion.
This is because we disrupt the equilibrium state by modulating the system. As the system energy gap is now smaller and more on resonance with the hot bath, additional energy can be transferred to the system before reaching a new equilibrium. Although the work output of the interaction modulation is improved, the system work output suffers slightly. This is due to the slight reduction in system energy when uncoupling the hot bath, as can be observed for the \ac{OLC} in \cref{fig:energy}. 
Overall, we get similar work output, but improved efficiency. This is due to the additional energy taken from the hot bath instead of the interaction work.  We conclude that considering the structure of the environment is indeed important when optimizing the cycle performance.

Using a different representation as in Fig.~\ref{fig:work_shifted}, it is instructive to discuss and compare the overall energy changes in terms of thermodynamic work diagrams (Fig.~\ref{fig:pv}, see also Ref.~\cite{KoyanagiNumericallyExactSimulations2022}). We stress that even though time no longer appears explicitly in these diagrams, the curves are simply derived from the time-dependent calculations as displayed in Figs.~\ref{fig:energy} and \ref{fig:work_shifted} and thus by no means we imply quasi-static state changes.

For the work diagrams, we regard the change of the Hamiltonian as a function of the externally controlled modulation functions \(f\) and \(h_{i}\).

From 
\begin{equation}
  \label{eq:work_contributions}
  \begin{aligned}
   \dd E {}&= \left\langle \frac{\partial H}{\partial t} \right\rangle \dd t = \ev{\partial_f H}\dd{f} + \sum_{i} \ev{\partial_{h_i}H} \dd{h_i} \\
         &=\left\langle \frac{\Omega (\sigma_z + 1)}{2} \right\rangle \dd f 
             + \sum_{i} \left\langle \frac{\sigma_x}{2}(B_i + B_i^\dagger) \right\rangle \dd h_i\\
         &\equiv - \dd W_S - \dd W_I,
  \end{aligned}
\end{equation}
we see that, e.g., the work due to the system modulation amounts to the area enclosed by the $f$-$\ev{\partial_f H}$-diagram.

In Fig.~\ref{fig:pv} we compare state changes using
the original \ac{OLC} protocol with the improved protocol where the cold bath is shifted by $\tau = 0.18\Theta$ in combination with slower transitions, i.e.\ $\tau_\mathrm{tr} = 0.12\Theta$.
The plots offer various insights.
First, it is evident that the work contribution due to the modulation of the interaction Hamiltonian must not be neglected in the overall energy balance, as the trajectories in the corresponding state space extend over similar regions.
Note that the individual subplots have been scaled such that work identified with the visual appearance of some area is indeed the same for all plots.
Second, although the shape of the system work diagram changes when the phases overlap (lower panel in Fig.~\ref{fig:pv}) compared to the \ac{OLC} (upper panel), the area and, thus, the work remains almost unaffected.
In contrast, the work due to the interaction terms changes significantly.
Most importantly, it increases for the hot as well as the cold bath contributions.
For the \ac{OLC} the positive work output due to the interaction with the hot bath is easily overtaken by the consumption due to the cold bath modulation (blue arrow in Fig.~\ref{fig:work_shifted}). On the other hand, the well-chosen overlap of the phases can maintain overall positive work output (green arrow in Fig.~\ref{fig:work_shifted}), leading to higher power and efficiency as explained earlier.

These results suggest that a systematic optimization of the coupling modulation speeds and timings should prove fruitful --
Bayesian optimization appears to be the method of choice, as the individual simulations are somewhat costly in computational resources.
Another interesting question for future study would be how the work extraction from the interaction modulation behaves if coherences in the work medium are present with the aim to make coherences less detrimental to the engine performance.

\section{Influence of cycle length and coupling strength}
\label{sec:infl-cycle-length}

In the previous section, the variation of the modulation protocol has been investigated.
Now we focus on how efficiency and power are affected by the overall length of the modulation cycle, and the coupling strength. Our results are in qualtitative agreement with Ref.~\cite{Wiedmann2020Mar}, where a similar investigation has been performed with a larger parameter range for an oscillator work medium. 

\begin{figure*}
\includegraphics[width=\textwidth]{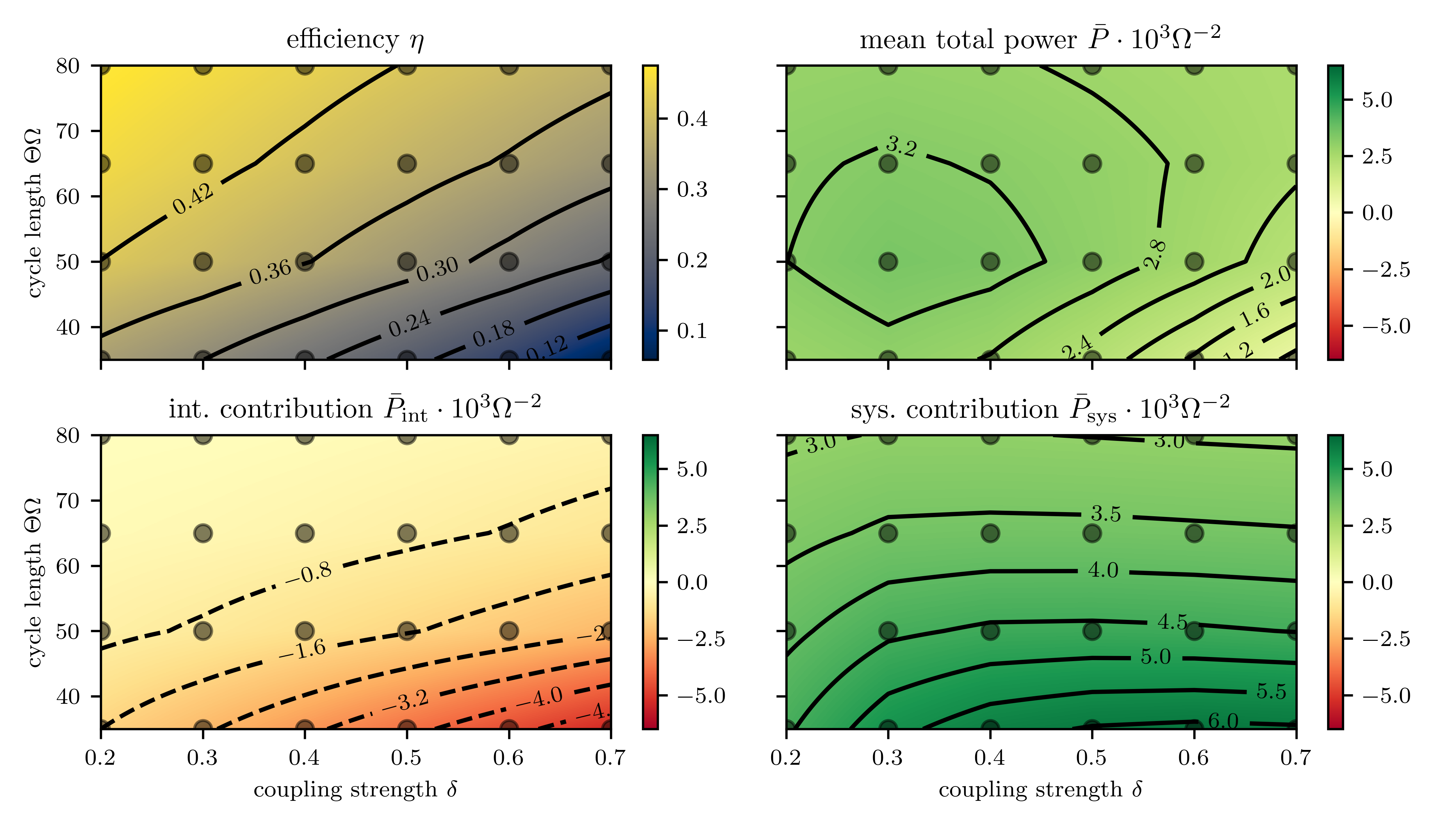}
    	\caption{
    	The average total power $\bar P$ and the efficiency $\eta$ are shown in the upper row.
    	The interaction and the system contribution which make up the total power are shown in the lower row.
    	At the gray dots, we have exact data from our simulation.
    	The values in between were obtained by linear interpolation.
    	As they serve as the basis for the contour lines, the plots should be considered as a rough estimate of the landscape while, still, showing the major features. These figures can be compared to Figs. 3, 4 and 5 of Ref.\ \cite{Wiedmann2020Mar} where qualitatively similar results have been found.
    	}
    	\label{fig:power_scan_linear}
\end{figure*}

While cycle duration can easily be adjusted from a practical perspective, tuning the coupling strength, especially beyond the weak coupling regime, is significantly more experimentally challenging \cite{Forn-DiazUltrastrongCouplingSingle2017}.
The results of our simulation for a rough scan over these two parameters are shown in Fig.~\ref{fig:power_scan_linear}.
In order to keep things as transparent as possible, the simulation was done for the \ac{OLC} modulation protocol without a $\sigma_x$ contribution in the system Hamiltonian (see also Fig.~\ref{fig:otto_like_cycle}).

As seen in the upper left panel of Fig.~\ref{fig:power_scan_linear}, the efficiency of the engine improves with weaker coupling and longer cycles, consistent with the findings of Ref\.~\cite{Wiedmann2020Mar}.
This agrees with the intuition that a weaker coupling amounts to a shallower slope of the coupling and decoupling from the environments.
The same holds true when increasing the cycle length, eventually approaching the quasistatic limit.
However, the power output (upper right panel in Fig.~\ref{fig:power_scan_linear}) is not trivially correlated with efficiency.
Instead, we find an optimal coupling strength of $\delta \approx 0.3$.
The decrease for weak couplings is evident. In the limit of zero coupling, no work can be extracted.
The decline for larger coupling strength is due to the larger losses associated with switching on and off the environments (bottom left panel of Fig.~\ref{fig:power_scan_linear}) while the system yield saturates (bottom right panel). This behavior agrees with the findings of Ref.~\cite{Wiedmann2020Mar}, where a maximum in the power output with respect to the coupling strength was observed, too.

For a fixed coupling strength, our results indicate that there is also an optimal cycle length with respect to the average power output.
It is evident that in the very slow driving regime (large $\Theta$, quasistatic regime) the average cycle power should be small and vanishes in the limit $\Theta \rightarrow \infty$.
The behavior for small $\Theta$, i.e.\ fast driving, is not as obvious.
Referring to the bottom right panel of Fig.~\ref{fig:power_scan_linear}, we see that the average system power increases while shortening the cycle.
This means that even though there is only little time for the system to be influenced by the baths, i.e.\ the system picks up/gives off only a small amount of energy, due to the shortening of the cycle, the power still increases.
It is sensible that for very short cycles, the work done by the system modulation scales linearly with the cycle length, thus leading to a finite power in the limit of $\Theta \rightarrow 0$.
On the other hand, a faster change of the system-bath interaction results in larger energy loss due to the coupling to the baths being switched on and off.
Summing up these two contributions yields a net power decrease when shortening the cycle.

\section{Methods}
\label{sec:method}

Our approach to determine the dynamics and to access bath observables of the microscopic open quantum system model \cref{eq:openSystemHamiltonian} relies on the stochastic description of the dynamics by means of the \ac{NMQSD} equation~\cite{Diosi1998Mar, Strunz1999Mar}, and its numerical treatment using
the \ac{HOPS}~\cite{Suess2014Oct, HartmannExactOpenQuantum2017, Hartmann2021Aug}.
For a comprehensive reading, we briefly review these methods before turning to the computation of bath observables in \cref{sec:bath-energy-change} and following sections.

\subsection{Non-Markovian quantum state diffusion (NMQSD)}
\label{sec:nmqsd}

In the usual model for open quantum systems~\cite{Caldeira1983Sep} each bath consists of an infinite set of harmonic oscillators and an interaction Hamiltonian that is linear in the ladder operators,
\begin{equation}
  \label{eq:generalmodel}
      H(t) =H_\sys(t) + \sum_{n=1}^N \Big( L_n^\dag(t)B_n + \hc \Big) + \sum_{n=1}^NH_B\nth ,
\end{equation}
with
\begin{equation}
  \begin{aligned}
    B_n={}&\sum_{\lambda} g_\lambda\nth a_\lambda\nth &
    H_B\nth={}&\sum_\lambda\omega_\lambda\nth {a^{\dag(n)}_\lambda} a_\lambda\nth.
  \end{aligned}
\end{equation}
The \(a_\lambda\nth\) are bosonic annihilation operators acting on the \(n\)th bath Hilbert space, $g_\lambda\nth$ denotes the individual coupling strengths, and \(L_n(t)\) is the system coupling agent. For NMQSD they can be considered arbitrary, time dependent, not necessarily Hermitian operators acting on the system Hilbert space, mediating the influence of the $n$th bath.
The bare system dynamics is governed by the (possibly time dependent) system Hamiltonian $H_\sys(t)$.
Quite general baths can be mapped to this model in the thermodynamic limit, when the bath consists of an infinite number of non-interacting sub-systems \cite{Makri1999Apr}.
Despite the simple harmonic structure of the bath, \cref{eq:generalmodel} is generally very hard to solve in a non-perturbative manner \cite{Breuer2002Jun}.
The \ac{NMQSD} approach tackles that problem by recasting \cref{eq:generalmodel} into a stochastic differential equation on the level of the system Hilbert space, in which the bath degrees of freedom are accounted for by Gaussian stochastic processes and a convolution term.
Note that in contrast to many of the perturbative approaches, the formalism is agnostic to a possible time dependence of the system Hamiltonian $H_\sys(t)$ and/or the coupling operators $L_n(t)$.

For the sake of simplicity, we review the \ac{NMQSD} (and the
\ac{HOPS}) method for a single bath -- the generalization to several baths is straightforward (see \cref{sec:hops_multibath}).
Furthermore, it is sufficient to treat the zero temperature initial condition only, i.e.\ a bath which is initialized in its ground state \(\ket{0}\). The effect of finite temperature can be accounted for correctly by an additional stochastic Hermitian contribution to the system Hamiltonian~\cite{HartmannExactOpenQuantum2017, Hartmann2021Aug}.

To derive the \ac{NMQSD} equation, the bath degrees of freedom are written in a basis of non-normalized Bargmann coherent states~\cite{klauder1968fundamentals}. These are given by \(|z_\lambda\rangle = \exp(z a_\lambda^\dagger)|0\rangle\) for a single bath mode.
The total system-bath state, thus, takes the form
\begin{equation}
  \label{eq:projected_single}
  \ket{ \Psi(t)} = \int{\frac{\dd[2]{\vb{z}}}{\bm{\pi}}\eu^{-\abs{\vb{z}}^2}}\ket{\psi(t,\vb{z}^\ast)}\ket{\vb{z}},
\end{equation}
where \(\vb{z}\) is the vector of all coherent state labels \(z_\lambda\), one label for each environmental oscillator.
The properties of Bargmann coherent states make the relative system states \(\ket{\psi(t,\vb{z}^{\ast})}\) a holomorphic function of \(\vb{z}^\ast\).

After transforming \cref{eq:generalmodel} into the interaction picture
with respect to \(H_\bath\) and using the properties of the coherent
states (\(\mel{z_\lambda}{a_\lambda}{\psi}\rightarrow \partial_{z_\lambda^\ast}\braket{z_\lambda}{\psi}\),
\(\mel{z_\lambda}{a_\lambda^\dag}{\psi}\rightarrow z_\lambda^\ast\braket{z_\lambda}{\psi}\)) we
deduce from the Schrödinger equation of the total state an equation for \(\ket{\psi(\vb{z}^{\ast},t)}\)
\begin{equation}
  \label{eq:nmqsd_single_proto}
    \iu \partial_t\ket{\psi(\vb{z}^{\ast},t)} =
    \begin{multlined}[t]
      \Biggl(H_\sys(t) + L(t) \sum_{\lambda}g_{\lambda}^\ast z_{\lambda}^\ast \eu^{\iu \omega_{\lambda} t}  \\
             + L^\dag(t) \sum_{\lambda} g_{\lambda}\eu^{-\iu \omega_{\lambda} t}\pdv{}{z_{\lambda}^\ast} \Biggr) \ket{\psi(\vb{z}^{\ast},t)}.
    \end{multlined}
\end{equation}
The key to eliminate the infinite sum is to shift perspective and regard \(\ket{\psi(t,\vb{z}^{\ast})}\) to be a functional of the Gaussian stochastic process
\begin{equation}
  \label{eq:single_process}
  \eta^\ast_{t} = -\iu \sum_\lambda g_\lambda^{\ast} z_\lambda^{\ast}\eu^{\iu \omega_\lambda t}.
\end{equation}
For a product initial condition $\ket{\psi_\sys(0)}\ket{0}_\bath$, this leads to the \ac{NMQSD} equation \cite{Diosi1998Mar, Strunz1999Mar}
\begin{equation}
  \label{eq:nmqsd_single}
  \begin{multlined}
   \partial_t\ket{\psi(\eta^\ast_t, t)} = (-\iu H_\sys(t) + L(t) {\eta}^\ast_t)\ket{\psi(\eta^\ast_t, t)}\\
  -L^\dag(t) \int_0^t\dd{s}\alpha(t-s)\fdv{\ket{\psi({\eta}^\ast_t, t)}}{\eta^\ast_s}\, .
  \end{multlined}
\end{equation}
When relating these pure states to the total system-bath state, we can read the integral in \cref{eq:projected_single} in a Monte-Carlo sense.
From the Gaussian distribution of the coherent state labels \(z_\lambda\) it follows that \(\eta_t\) is a Gaussian stochastic process which obeys
\begin{equation}
  \label{eq:single_processescorr}
  \begin{gathered}
    \mathcal{M}(\eta_t) =0,\quad   \mathcal{M}(\eta_t\eta_s) = 0, \\
    \mathcal{M}(\eta_t\eta_s^\ast) = \alpha(t-s)\, .
  \end{gathered}
\end{equation}
This shows that in the \ac{NMQSD} formalism the microscopic structure of the bath, i.e.\ the values of the individual coupling strengths \(g_\lambda\), is fully accounted for in terms of the zero temperature \ac{BCF}
\begin{equation}
  \label{eq:bcfdef}
  \alpha(t) = \sum_\lambda \abs{g_\lambda}^2\eu^{-\iu \omega_\lambda t} = \frac{1}{\pi}  \int\dd \omega J(\omega) \eu^{-\iu \omega t}\, ,
\end{equation}
as introduced earlier in (\ref{bcf}).
In the continuous limit 
(infinitely many bath oscillators), the spectral density becomes a continuous function of $\omega$ and the \ac{BCF} decays to zero for \(\tau\rightarrow \infty\).
Note that \cref{eq:nmqsd_single} reveals that the propagation from $0$ to $t$ requires knowledge of the bath correlation function from $0$ to $t$, only.

Tracing out the bath degrees of freedom amounts to averaging the dyadic product of stochastic pure states for many realizations of the stochastic process and, thus, the reduced state reads
\begin{equation}
  \label{eq:recover_rho}
    \rho_{\sys}(t) = \mathcal{M}_{\eta_{t}^\ast}\big(\ketbra{\psi(\eta_t, t)}{\psi(\eta^\ast_t, t)}\big) \,.
\end{equation}

Notably, \cref{eq:nmqsd_single} does not preserve the norm of the state vector, leading to flawed stochastic convergence of \cref{eq:recover_rho}.
This can be overcome within the nonlinear \ac{NMQSD} formalism~\cite{Diosi1998Mar} which naturally implements importance sampling of the trajectories.
The formal expression
\begin{equation}
  \label{eq:norm_av}
  \begin{multlined}
          \rho_{\sys}(t) =
 \int{\frac{\dd[2]{\vb{z}}}{\bm{\pi}}\Bigg[\eu^{-\abs{\vb{z}}^2}}
 \braket{\psi(t,\vb{z})}{\psi(t,\vb{z}^\ast)}\\\times \frac{\ketbra{\psi(t,\vb{z})}{\psi(t,\vb{z}^\ast)}}{\braket{\psi(t,\vb{z})}{\psi(t,\vb{z}^\ast)}}\Bigg].
  \end{multlined}
\end{equation}
is then seen as an average over normalized dyads, i.e.\ a density matrix of a pure state, with a time-dependent weight function $\eu^{-\abs{\vb{z}}^2}\braket{\psi(t,\vb{z})}{\psi(t,\vb{z}^\ast)}$ which amounts to the Husimi  \(Q\) function of the bath~\cite{Diosi1998Mar,Suess2014Oct,RichardDiss}.
The sampling according to such a weight is achieved by using the shifted stochastic process
\begin{equation}
  \label{eq:shifted_proc}
  \tilde{\eta}_{t}^\ast= \eta^\ast_{t} + \int_{0}^{t}\dd{s} \alpha^\ast(t-s) \ev{L^\dagger}_{s}
\end{equation}
with
\begin{equation}
    \ev{L^\dag}_{t}=\frac{\mel{\psi(\tilde{\eta}_{t}, t)}{L^\dag}{\psi(\tilde{\eta}_{t}^\ast,t)}}{\langle\psi(\tilde{\eta}_{t}, t)|\psi(\tilde{\eta}_{t}^\ast,t)\rangle}
\end{equation}
and a propagation of the pure state trajectories with the nonlinear \ac{NMQSD} equation
\begin{equation}
  \label{eq:nmqsd_nonlin_single}
  \begin{multlined}
  \partial_t\ket{\psi(\tilde{\eta}^\ast_t, t)} = (-\iu H_\sys + L {\tilde{\eta}}^\ast_t)\ket{\psi(\tilde{\eta}^\ast_t, t)} 
  \\- \pqty{L^\dag -\ev{L^\dag}_{t}}\int_0^t\dd{s}\alpha(t-s)\fdv{\ket{\psi({\tilde{\eta}}^\ast_t, t)}}{\tilde{\eta}^\ast_s}.
  \end{multlined}
\end{equation}

From the solutions of the nonlinear \ac{NMQSD} \cref{eq:nmqsd_nonlin_single}, the reduced system state is recovered through
\begin{equation}
  \label{eq:recover_rho_nonlinear}
  \rho_{\sys}(t) =
  \mathcal{M}_{\tilde{\eta}_{t}^\ast}\qty(\frac{\ketbra{\psi(\tilde{\eta}_t, t)}{\psi(\tilde{\eta}^\ast_t,t)}}{\braket{\psi(\tilde{\eta}_t, t)}{\psi(\tilde{\eta}^\ast_t,t)}}),
\end{equation}
so that all trajectories contribute with ``equal weight''.

\subsection{Hierarchy of stochastic pure states (HOPS)}
\label{sec:hops}
In equation \eqref{eq:nmqsd_single} the bath degrees of
freedom have been removed from explicit consideration, replacing them with a Gaussian
stochastic process and a rather complicated term containing a convolution
integral with a functional derivative
\begin{equation}
  \label{eq:complicated_term}
  D(t) := \int_0^t\dd{s}\alpha(t-s)\fdv{\ket{\psi({\eta}^\ast_t)}}{\eta^\ast_s}.
\end{equation}
In special cases, there are analytical and perturbative expressions for this
term~\cite{Diosi1998Mar}, but we keep the approach as
general as possible and instead choose a numerical avenue.

The key to proceed is to cast the complicated term containing the functional derivative into
a set of auxiliary states~\cite{Suess2014Oct,HartmannExactOpenQuantum2017,Hartmann2021Aug,RichardDiss}.
As is done in the \ac{HEOM} approach, we expand the \ac{BCF} into exponentials
\(\alpha(\tau)=\sum_{\mu=1}^M G_{\mu}\eu^{-W_{\mu}\tau}\) with $G_{\mu}, W_{\mu} \in \mathbb{C}$. With
\begin{equation}
  \label{eq:d_op_one}
      D_{\mu}(t) \equiv \int_{0}^{t}\dd{s} G_{\mu} \eu^{-W_{\mu}(t-s)} \fdv{\eta^\ast_s} 
\end{equation}
and
\begin{equation}
  \label{eq:d_op_one_full}
  D^{\vb{k}}(t) \equiv \prod_{\mu=1}^{M}\sqrt{\frac{1}{G_{\mu}^{k_{\mu}}k_{\mu}!}} \frac{1}{\iu^{k_{\mu}}}\pqty{D_{\mu}(t)}^{k_{\mu}}
\end{equation}
we can define the \(\vb{k}th\) hierarchy state
\begin{equation}
  \label{eq:d_op_hier}
  \ket{\psi^{\vb{k}}}\equiv
  \begin{cases}
    D^{\vb{k}}\ket{\psi} &  \text{for } k_{\mu} \geq 0\, \forall \mu\\
    0 & \text{otherwise.}
  \end{cases}
\end{equation}
The origin of the normalization chosen in \cref{eq:d_op_hier} is the
desire to give all hierarchy states the same unit and to formulate the
final \ac{HOPS} equations unified into one equation in an enlarged Hilbert
space as is done in~\refcite{Gao2021Sep}. This is a very convenient choice, but other choices of normalization may be made. 

For a hierarchy state, the following equation of motion can be
derived~\cite{Suess2014Oct,Hartmann2021Aug}
\begin{widetext}
  \begin{equation}
  \label{eq:singlehops}
  \begin{aligned}
  \partial_t \ket{\psi^{\vb{k}}} 
  = {}& \Big(-\iu H_\sys(t) + L(t) \eta^\ast(t) - \sum_{\mu=1}^{M}k_{\mu}W_\mu\Big)\ket{\psi^{\vb{k}}} - \iu \sum_{\mu=1}^{M}\sqrt{G_\mu k_{\mu}}           L(t)   \ket{\psi^{\vb{k} - \vb{e}_{\mu}}} 
      - \iu \sum_{\mu=1}^{M}\sqrt{G_\mu\qty(k_{\mu} + 1)}  L^\dag(t) \ket{\psi^{\vb{k} + \vb{e}_{\mu}}} \,,
  \end{aligned}
\end{equation}
\end{widetext}
where \(\vb{k}=(k_{1}, k_{2}, \ldots, k_{M})\) with \(k_{\mu}\geq 0\) is a multi index and \(\pqty{\vb{e}_{\mu}}_{\nu} = \delta_{\mu,\nu}\). The norm of the multi index \(\vb{k}\) is denoted \(\abs{\vb{k}}=\sum_{\mu}k_{\mu}\). We call this norm the hierarchy level of \(\ket{\psi^{\vb{k}}}\).  The zero-level hierarchy state corresponds to the pure state trajectory obeying \cref{eq:nmqsd_single}, i.e, \(\ket{\psi(\eta_{t}^\ast, t)}\equiv \ket{\psi^{\vb{0}}}\).  Since the \ac{HOPS} does no longer involve the functional derivative, we omit the explicit dependency on the stochastic process in the notation for the elements $\ket{\psi}^{\vb{k}}$ of the composite hierarchy state $(\dots, \ket{\psi}^{\vb{k}}, \dots)$.  \cref{eq:singlehops} is the stochastic \emph{Hierarchy of Pure States} (HOPS) -- each state couples to states of neighboring level.  This is similar to the \ac{HEOM} approach~\cite{Kato2016Dec}, but with the advantage of reducing the dimensionality of the auxiliary objects from \(\dim{\hilb_{\sys}}^{2}\) to \(\dim{\hilb_{\sys}}\) by treating pure states instead of density matrices.

It has been shown that the \ac{HOPS} for the nonlinear \ac{NMQSD} equation is obtained by, again, replacing \(\eta\rightarrow \tilde{\eta}\) and \(L^\dag\rightarrow \pqty{L^\dag-\ev{L^\dag}_{t}}\)~\cite{Suess2014Oct, HartmannExactOpenQuantum2017, Hartmann2021Aug}.  When truncating the hierarchy, the resulting finite system of nonlinear differential equations with time dependent coefficients can be solved numerically using standard ODE-solvers.  By increasing the cutoff level, the numerical error can be made arbitrarily small~\cite{Hartmann2021Aug}. There are various truncation schemes available~\cite{RichardDiss,Zhang2018Apr,Hartmann2021Aug}. As we are not resource limited in this work, a simple simplex truncation condition was used, i.e.\ \(\abs{\vb{k}} \leq k_{\max}\).  
We show in the next section that they can be of great use beyond the mere calculation of the root stochastic pure state \(\ket{\psi(\tilde{\eta}^\ast_t,t)}=\ket{\psi^{\vb{0}}}\).

The \ac{HOPS} formalism presented here relies on an exponential representation of the \ac{BCF}.
Finding such multi-exponential representations is a common problem in the field of open quantum systems~\cite{MeierNonMarkovianEvolutionDensity1999, RitschelAnalyticRepresentationsBath2014a, XuTamingQuantumNoise2022}.
For an Ohmic \ac{SD} with exponential cutoff (as used in this work, see \cref{eq:ohmic_sd})
\begin{equation}
  J(\omega) \sim \omega \eu^{-\omega/\omega_c}
\end{equation}
the resulting \ac{BCF} decays algebraically. Thus,
any finite sum of exponential terms will only ever serve as an approximation.
However, by numerically fitting the exponential expression to the \ac{BCF}, the algebraic decay is well reproduced over several orders of magnitude using only a few exponential terms~\cite{HartmannExactOpenQuantum2017, RichardDiss}.
This allows us to faithfully propagate the open quantum system efficiently.

\subsection{Bath energy change}
\label{sec:bath-energy-change}

\subsubsection{Preliminaries}
\label{sec:bath-energy-change-pre}
After reviewing the fundamentals of the \ac{NMQSD} equation and the \ac{HOPS} approach in \cref{sec:nmqsd,sec:hops}, we are now in a position to calculate the change of the bath energies -- a novel theoretical aspect within the \ac{NMQSD}/\ac{HOPS} framework.

As elucidated before, the \ac{NMQSD} equation determines the dynamics of the stochastic trajectory
\(\ket{\psi(\eta^\ast_t, t)}\), which is a projection of the
global state of system and bath onto a coherent bath state. In this
way, the complete information about the global state is encoded in the
ensemble of trajectories. Accessing individual properties of a single bath
oscillator would be challenging, since the influence of the
bath as whole is mediated and encoded through the collective stochastic process. Still, we can expect to be
able to access collective bath quantities, such as 
\(B_n=\sum_{\lambda} g_\lambda\nth a_\lambda\nth\), i.e.\ a part of the interaction
Hamiltonian of the $n$th bath in \cref{eq:generalmodel}. 

In order to demonstrate and elaborate further this general idea, we determine the change of the bath energy
\begin{equation}
  \label{eq:heatflowdef}
  J = - \dv{\ev{H_\bath}}{t}\, .
\end{equation}
The choice of the negative sign ensures $J$ is attributed with the flow of energy out of the bath.
For the time being, we stick to the simplest version of the general model \cref{eq:generalmodel} with a single bath at zero temperature and no explicit time dependence of the Hamiltonian
\begin{equation}
  \label{eq:totalH}
  H = H_\sys + \underbrace{LB^\dag + L^\dag B}_{H_\inter} + H_\bath \, .
\end{equation}
While the bath energy expectation value \(\ev{H_{\bath}}\) is generally infinite, \(\partial_{t}\ev{H_{\bath}}\) will remain finite. 

From the Ehrenfest theorem, it follows
\begin{equation}
  \label{eq:ehrenfest}
  \i\partial_t\ev{H_\bath} = \ev{[H_\bath,H]} = \ev{[H_\bath,H_\inter]}
\end{equation}
with \([H_\bath, B^\dag ]=\sum_\lambda \omega_\lambda g^\ast_\lambda a^\dag_\lambda\).
Noting that $B^\dag$ in the interaction picture with respect to \(H_\bath\) reads $B^\dag(t) = \sum_\lambda g^\ast_\lambda a^\dagger_\lambda \eu^{\iu \omega_\lambda t}$ we find
\begin{equation}
  \label{eq:expcomm}
    \ev{[H_\bath,H_\inter]} = \frac{1}{\i}\qty(\ev{L\dot{B}^\dag}_\inter  + \cc)
\end{equation}
where we switched to the interaction picture with respect to \(H_\bath\) to
also connect to the \ac{NMQSD} formalism.
Thus, the energy flow out of the bath is
\begin{equation}
  \label{eq:final_flow}
  J(t) = \ev{L^\dag \dot{B}(t) + L\dot{B}^\dag(t)}_\inter.
\end{equation}
We will remain in the interaction picture and drop
the subscript in the following.

Notably, the expectation value \(\ev{L^\dag\partial_t B(t)}\) can be connected to the first-level auxiliary states of the \ac{HOPS}.
As for the derivation of the \ac{NMQSD} equation (\cref{sec:nmqsd}), using non-normalized Bargmann
coherent states $\ket{\vb{z}}$ we write
\begin{equation}
  \label{eq:interactev}
      \ev{L^\dag\partial_t B(t)} = \int\frac{\dd[2]{\vb{z}}}{\bm{\pi}} \braket{\Psi(t)}{\vb{z}}\mel{\vb{z}}{L^\dag\partial_tB(t)}{\Psi(t)}
\end{equation}
and find
\begin{equation}
  \label{eq:nmqsdficate}
      \begin{multlined}[t]
  \mel{\vb{z}}{\partial_tB(t)}{\psi(t)}
    {}= \sum_\lambda g_\lambda
  \qty(\partial_t \eu^{-\i\omega_\lambda
    t})\partial_{z^\ast_\lambda}\ket{\psi(z^\ast,t)} \\
 \begin{aligned}
      {}&= \int_0^t \sum_\lambda g_\lambda
  \qty(\partial_t \eu^{-\i\omega_\lambda
    t})\pdv{\eta_s^\ast}{z^\ast_\lambda}\fdv{\ket{\psi(z^\ast,t)}}{\eta^\ast_s}\dd{s}\\
  {}&= -\i\int_0^t\dot{\alpha}(t-s)\fdv{\ket{\psi(\eta^\ast_{t},t)}}{\eta^\ast_s}\dd{s},
 \end{aligned}
  \end{multlined}
\end{equation}
which yields
\begin{equation}
  \label{eq:steptoproc}
  \begin{multlined}
      \ev{L^\dag\partial_t B(t)} =\\
        -\i \mathcal{M}_{\eta^\ast_t}\Big[\bra{\psi(\eta_{t}, t)}
         L^\dag\int_0^t\dd{s} \dot{\alpha}(t-s)\fdv{\eta^\ast_s}
  \ket{\psi(\eta^\ast_{t},t)}\Big]\,,
    \end{multlined}
\end{equation}
an expression that only depends on the stochastic trajectory and the \ac{BCF}.

Within the \ac{HOPS} formalism (\cref{sec:hops}), where we have assumed
\(\alpha(t) = \sum_\mu G_\mu \eu^{-W_\mu t}\), the energy flow can thus be expressed in terms of an ensemble average involving the first-level hierarchy states \(\ket{\psi^{\vb{e}_\mu}(t)}\),
\begin{equation}
  \label{eq:hopsflowfock}
  \begin{multlined}
    J(t) =
    - \sum_\mu \Big(\sqrt{G_\mu}W_\mu
    \mathcal{M}_{\eta^\ast} \mel{\psi^{\vb{0}}(t)}{L^\dag}{\psi^{\vb{e}_\mu}(t)}\Big) + \cc
  \end{multlined}
\end{equation}

\subsubsection{Finite temperature}
\label{sec:non-zero_temperature}

In the general \ac{NMQSD}/\ac{HOPS} formalism, finite temperature can be included in many different ways \cite{Diosi1998Mar,Strunz1999Mar,deVega2005,Strunz2005,deVega2017}. Using \ac{HOPS}, it is most convenient to include finite temperature through a stochastic (Hermitian) contribution to the system Hamiltonian~\cite{HartmannExactOpenQuantum2017, Hartmann2021Aug}: for an arbitrary observable $O$, not necessarily of the system, its expectation value at time $t$
can be written as
\begin{equation}
    \ev{O}(t) = \tr \left[O U(t) (\rho_\sys\otimes \rho_\bath) U^\dagger(t) \right]
\end{equation}
where $U(t)$ denotes the global unitary time evolution operator and $\rho_\sys \otimes \rho_\bath$ is the factorized system-bath initial condition.
For a Gibbs state \(\rho_\bath={\eu^{-\beta H_\bath}}/{Z}\) in its
Glauber-Sudarshan P representation we may write
\begin{equation}
  \begin{multlined}
        \langle O \rangle(t) =
    \prod_\lambda \qty(\int\dd[2]{y_\lambda}
    \frac{\eu^{-\abs{y_\lambda}^2 / \bose_\lambda}}{\pi\bose_\lambda})
  \\\times \tr[O U(t) \rho_\sys D(\vb{y})\ketbra{0}D^\dag(\vb{y}) U^\dag(t)],
  \end{multlined}
\end{equation}
where the coherent states have already been expressed in terms of the multimode shift operator
\begin{equation}
  \label{eq:shiftop}
  \begin{aligned}
    D(\vb{y}) &= \bigotimes_\lambda \eu^{y_\lambda a_\lambda^\dag-y^\ast_\lambda a_\lambda},&  D(\vb{y})\ket{0} &= \ket{\vb{y}} .
  \end{aligned}
\end{equation}
Further, $\bose_{\lambda}$ denotes the Bose-Einstein distribution \(\bose_{\lambda}=\bose(\beta\omega_{\lambda}) = (\eu^{\beta\omega_\lambda}-1)^{-1}\).
Using the unitarity of $D(\vb{y})$, cyclic permutations under the trace and reading the $y_\lambda$-integrals in a Monte-Carlo sense as an average over the Gaussian \(\vb{y}\)-coherent state labels results in
\begin{equation}
    \langle O \rangle(t) = \mathcal{M}_{\vb{y}} \tr \tilde O(\vb{y}) \tilde U(\vb{y},t) \rho_\sys \ketbra{0} \tilde U^\dag(\vb{y}, t) \, .
\end{equation}
This expression involves the propagation of an initial product state where the bath is now in its ground state and can, thus, be calculated using the zero-temperature-\ac{HOPS} introduced earlier. Note, however, that the time evolution is determined by the stochastically shifted Hamiltonian $\tilde H(\vb{y}, t) = D^\dagger(\vb{y}) H(t) D(\vb{y})$. In our case,
referring to the \ac{NMQSD} formalism (open system Hamiltonian \eqref{eq:generalmodel} in the interaction picture with respect to $H_\bath$), the transformation simply adds to the original Hamiltonian $H(t)$ a stochastic system contribution
\begin{equation}
  \label{eq:thermalh}
  H_{\sys}^{\mathrm{shift}}(\xi, t) = L \xi^{*}(t)+L^{\dag} \xi(t)\,.
\end{equation}
The function
\begin{equation}
  \label{eq:xiproc}
  \xi(t)= \sum_{\lambda} g_{\lambda} y_{\lambda} \eu^{-\mathrm{i} \omega_{\lambda} t}
\end{equation}
depends on the values of $\vb{y}$, which are complex valued Gaussian random variables with
$\mathcal{M}(y_{\lambda}) = 0 = \mathcal{M}(y_{\lambda}y_{\lambda'})$ and
$\mathcal{M}(y_{\lambda}y^\ast_{\lambda'}) = \bose_\lambda \delta_{\lambda, \lambda'}$.
Thus, $\xi(t)$ can be seen as a Gaussian stochastic process completely specified through its moments
\(\mathcal{M}(\xi(t))=0=\mathcal{M}(\xi(t) \xi(s))\) and
\begin{equation}
    \mathcal{M}\left(\xi(t) \xi^{*}(s)\right)
    =\frac{1}{\pi} \int_{0}^{\infty} \dd{\omega} \bose(\beta \omega) J(\omega) \eu^{-\iu \omega(t-s)} \,,
\end{equation}
closely related to the thermal part of the usual bath correlation function.
Specifically, for the change of the bath energy we have $O=L^\dagger \partial_t B(t)$.
Its transformed expression becomes
\begin{equation}
  \begin{multlined}
    \tilde O(\vb{y}) = D^\dagger(\vb{y}) O D(\vb{y})
    = L^\dagger \partial_t (B(t) + \xi(t)) = \tilde O(\xi)\, .
  \end{multlined}
\end{equation}
It follows that for a non-zero temperature initial condition we obtain
\begin{equation}
    \ev{L^\dagger \partial_t B(t)} = \mathcal{M}_\xi \ev{L^\dagger \partial_t (B(t) + \xi(t))}_\xi
\end{equation}
where the expectation value on the right-hand side has to be taken with respect to states
that have been propagated with a zero-temperature initial condition and a system Hamiltonian with
additional stochastic shift $H_\sys^{\mathrm{shift}}(\xi, t)$.
Finally, within the \ac{HOPS} formalism the change of the bath energy takes the form
\begin{equation}
  \label{eq:bath_energy_change_with_temp}
  J(t) =- \mathcal{M}_{\eta^\ast, \xi}
    \begin{aligned}[t]
      \Bigg( &\sum_\mu\sqrt{G_\mu}W_\mu \bra{\psi^{\vb{0}}(t)}L^\dag  \ket{\psi^{\vb{e}_\mu}(t)}\\
  &+ \bra{\psi^{\vb{0}}(t)}L^\dag \dot \xi(t)\ket{\psi^{\vb{0}}(t)} \Bigg)+ \cc \,
    \end{aligned}
\end{equation}
The appearance of the derivative \(\dot{\xi}(t)\) of the stochastic process may look troublesome.
However, if the auto-correlation function \(\mathcal{M}(\xi(t)\xi^\ast(s))\) is twice differentiable, as in our case, the sample trajectories are smooth.
When sampling such trajectories using Fourier-methods~\cite{RichardDiss, HartmannStocproc2023}, their time derivative can be calculated equally well.

\subsubsection{N baths in the nonlinear formalism}
\label{sec:generalizations}

Referring to the general Hamiltonian \eqref{eq:generalmodel} with $N$ baths,
the \ac{NMQSD}/\ac{HOPS} formalism can be generalized with no major difficulties (see \cref{sec:hops_multibath}).
From the $N$ \acp{BCF} $\alpha_n(\tau)$ one deduces $N$ stochastic processes $\eta^\ast_n(t)$ and
their thermal correspondent $\xi_n(t)$.
Further, the hierarchy index vector $\vb{k}$ generalizes to $N$ such vectors $\kmat = (\vb{k_1}, \dots \vb{k_N})$.
Employing the nonlinear \ac{NMQSD} equation to benefit from the superior stochastic convergence, we arrive at the following expression for the change of the energy of the $n$th bath
\begin{widetext}
\begin{equation}
      J_n(t) = - \mathcal{M} \Bigg(
      \sum_\mu\sqrt{G_\mu^{(n)}}W_\mu^{(n)} \frac{\bra{\psi^{\mat{0}}(t)}L_n^\dag \ket{\psi^{\mat{e}_{n,\mu}}(t)}}{\bra{\psi^{\mat{0}}(t)}\ket{\psi^{\mat{0}}(t)}}
      + \frac{\bra{\psi^{\mat{0}}(t)}L_n^\dag \dot \xi_n(t)\ket{\psi^{\mat{0}}(t)}}{\bra{\psi^{\mat{0}}(t)}\ket{\psi^{\mat{0}}(t)}}\Bigg) + \cc \, .
\end{equation}
\end{widetext}
The average $\mathcal{M}$ is taken with respect to the $2N$ stochastic processes $(\eta^\ast_1, \dots \eta^\ast_N, \xi_1, \dots \xi_N$).
The \ac{HOPS} vector $(\dots, \ket{\psi}^{\mat{k}}, \dots)$ is propagated using the nonlinear \ac{HOPS} (see \cref{sec:nmqsd,sec:hops}).

\subsection{Interaction energy}
\label{sec:interaction_energy}

The interaction energy between the system and the $n$th bath is directly accessible by the \ac{HOPS} formalism, too.
From the expression
\begin{equation}
    \ev{H_\inter^{(n)}} = \ev{L B^{\dagger(n)} (t) + L^\dagger B^{(n)}(t)}
\end{equation}
and the derivation shown in \cref{sec:bath-energy-change}, one immediately finds
\begin{equation}
  \label{eq:interaction_energy_hops}
  \begin{multlined}
    \ev{H_\inter^{(n)}} =\mathcal{M} \Bigg(
    \sum_\mu\sqrt{G_\mu^{(n)}} \frac{\bra{\psi^{\mat{0}}(t)}L_n^\dag \ket{\psi^{\mat{e}_{n,\mu}}(t)}}{\bra{\psi^{\mat{0}}(t)}\ket{\psi^{\mat{0}}(t)}}
    \\+ \frac{\bra{\psi^{\mat{0}}(t)}L_n^\dag \xi_n(t)\ket{\psi^{\mat{0}}(t)}}{\bra{\psi^{\mat{0}}(t)}\ket{\psi^{\mat{0}}(t)}}\Bigg) + \cc \, ,
  \end{multlined}
\end{equation}
where, again, the average $\mathcal{M}$ is taken with respect to $(\eta^\ast_1, \dots \eta^\ast_N, \xi_1, \dots \xi_N$) and nonlinear \ac{HOPS} determines the evolution of $\ket{\psi^{\mat{k}}(t)}$.

\subsection{Total power}

Since the derivations in the previous sections did not include any
explicit time derivatives of the system Hamiltonian $H_\sys$ and the coupling
operators $L_n$, we can simply substitute \(H_\sys\rightarrow H_\sys(t)\) and
\(L\rightarrow L(t)\) in any of the expressions.
In case of such explicit time-dependencies of the open system Hamiltonian
\eqref{eq:openSystemHamiltonian}, the total power
\begin{equation}
  \label{eq:total_power}
  \dv{\ev{H}}{t} = \ev{\pdv{H_\inter}{t}} + \ev{\pdv{H_\sys}{t}}
\end{equation}
will in general be non-zero.
It can be calculated using the \ac{HOPS} by simply replacing \(L(t)\) with \(\dot{L}(t)\) in \cref{eq:interaction_energy_hops}. 

\subsection{Other collective bath observables}

Finally, for the record, 
we note that the dynamics of any observable that can be expanded in
terms of the collective bath operators \(B_n\) and \(B_n^\dagger\) of \cref{eq:generalmodel} can be calculated using the \ac{HOPS} formalism.
It turns out that the expectation value of an operator $O$
with normal ordering in $B_n$ and $B_n^\dagger$, i.e.\
\begin{equation}
  O = F \otimes \bqty{B_i^\dag}^{l}\bqty{B_j}^{m}
\end{equation}
where $F$ acts on the system only, can be expressed in terms of
auxiliary states of level $l$ and $m$.
Details are given in \cref{sec:gener-coll-bath}.

The expectation value of the interaction energy  discussed above is
a particular example of this kind.

\section{Conclusion and outlook}
\label{sec:conclusion}

In this article, we expand the numerically exact \ac{HOPS} method and apply it to the dynamics of a quantum heat engine. Both the work medium and bath degrees of freedom are explicitly included.
Since the approach allows for arbitrary time-dependent couplings and system Hamiltonians, we are able to tackle realistic quantum thermodynamic cycles. 
The \ac{HOPS} has originally been developed for the numerically exact description of the reduced dynamics of open quantum systems. 
In this paper, we extend the method to also access the dynamics of relevant bath observables. 
In particular, we are able to follow the various energy contributions in time, i.e.\ the system, interaction, and bath energies.
On such a global scope it is then straightforward to define power and efficiency based on well-defined energy expectation values, circumventing ambiguities that may arise when working on a reduced level (e.g.\ with master equations).
We find that the energetics of a quantum heat engine can be highly sensitive to the details of the modulations during the cycle. 
Even slight changes in the protocol can significantly affect the power and efficiency of the machine. For instance, overlapping phases can lead to increased power output and efficiency.

We illustrate our approach to quantum heat engines by taking a closer look at the Otto cycle with a qubit as work medium and two Ohmic baths. 
Special emphasis is placed on the per-phase relaxation dynamics of the qubit, showing once again that the steady state does not necessarily correspond to the Gibbs ensemble of the instantaneous system Hamiltonian.
Furthermore, a detailed discussion of the behavior of the individual energy contributions of the global Hamiltonian with a fine resolution on the phases of the cycle is given.
The loss in overall energy quantifies the extracted work and yields the cycle averaged power.
Viewing the energy loss from the hot bath as the energy needed to drive the cycle, the efficiency is calculated rigorously.
The dependence of power and efficiency on the parameters of the model is addressed carefully, investigating a broad range.
We find that when keeping the overall coupling strength and cycle length fixed, altering the speed and timing of the modulation protocol can yield significant enhancements.
Notably, for a particular overlap between the hot and the cold phases, we discover that the work contribution due to the modulation of the interaction Hamiltonian becomes positive.
Thus, the process of coupling to and decoupling from the baths itself contributes positively to the amount of work done by the engine.
Using thermodynamic work diagrams, a helpful visualization of these findings is possible.
Finally, we discuss in detail how the coupling strength and the cycle length influence power and efficiency.
As expected, the efficiency increases when moving towards the quasistatic case.
For the total power, which is composed of competing contributions from the system and the interaction part, we find that there is an optimal choice for both parameters.
In general, our thorough analysis, enabled by the \ac{HOPS} method, reveals that thermodynamic properties of a quantum heat engine are very sensitive to the details of a fully quantum dynamical treatment, which are often lost by a too simplistic reduced description.
Finally, we want to emphasize that the \ac{HOPS} is equally applicable to other work media, non-Ohmic spectral densities, arbitrarily low temperatures, and any time-dependent system Hamiltonian and coupling, making it a universal method for describing general quantum dynamical (thermodynamic) scenarios.

For concrete applications, the modulation protocol with the highest efficiency and the largest power has yet to be found, depending on specific experimental parameters. Our investigations, however, show that such an optimized protocol will most likely include overlapping phases. 
Also, instead of introducing a semi-classical external modulation, autonomous heat engines with an explicit quantum energy storage would be of interest in our HOPS framework~\cite{Binder2018}. 
Such a more explicit inclusion of the work extraction process would also be necessary to answer the question of how the work generated by such a single-qubit heat engine can actually be used or measured.
All thermodynamic quantities presented in this article have to be understood in the usual sense of quantum expectation values. 
It is still an open question, and the subject of ongoing research, to what extent the work extracted from single-shot experiments is measurable and deterministically usable at all~\cite{frenzelReexaminationPureQubit2014a,horodeckiFundamentalLimitationsQuantum2013,niedenzuConceptsWorkAutonomous2019,beyerMeasurementQuantumWork2023,dahlstenInadequacyNeumannEntropy2011b,stevensEnergeticsSingleQubit2022}. 
On a more fundamental level, we are confident that the possibility to access bath observables will allow for further insights in the connection between strongly coupled open quantum system dynamics and thermodynamics.  
With the help of the extended HOPS method, recent proposals for entropic quantities and  generalizations of the thermodynamic laws at strong coupling can be tested for specific, realistic model systems far from the weak coupling limit
\cite{strasbergNonequilibriumThermodynamicsStrong2016,perarnau-llobetStrongCouplingCorrections2018,hsiangQuantumThermodynamicsStrong2018,talknerColloquiumStatisticalMechanics2020,sakamotoNumericallyExactSimulations2020a,rivasStrongCouplingThermodynamics2020,collaOpensystemApproachNonequilibrium2022,alipourEntropybasedFormulationThermodynamics2022,anto-sztrikacsEffectiveHamiltonianTheoryOpen2023}.

Further improvements of the method are desirable and are subject of ongoing research. 
In particular, depending on the details of the model, reaching a limit cycle in the HOPS simulations can be computationally expensive and the initial, transient dynamics is of little interest with respect to the thermodynamic properties of the machine.
Therefore, a more direct access to the steady-state behavior of the heat engine would be valuable. The modulation protocol in this work was still slow enough so that the work medium could reach a steady state in most of the phases. It remains to conduct a more detailed study for faster protocols with off-diagonal system modulations to explore the effects of coherences in the work medium, especially with respect to the shifted (overlapping) protocols.

Finally, it would be fascinating to investigate many-body work media using the methods suggested in this work.

\section*{Acknowledgements}
 We thank Valentin Link and Kimmo Luoma for valuable discussions and helpful suggestions. WTS wants to thank Yoshitaka Tanimura for insightful conversations during a Freiburg workshop. 
 All authors are grateful to the Centre for Information Services and High Performance Computing [Zentrum für Informationsdienste und Hochleistungsrechnen (ZIH)] at TU Dresden for providing its facilities for high throughput calculations.

\section*{Data availability statement}
The data that support the findings of
this study are available from the
corresponding author upon reasonable
request. For the source code of the method used in this work and notes on the numerics see~\cref{sec:notes-numerics}.
\section*{References}
\bibliographystyle{unsrt}
\bibliography{index}

\appendix{}

\section{NMQSD and HOPS for multiple baths}
\label{sec:hops_multibath}

We generalize the Non-Markovian Quantum State Diffusion (\ac{NMQSD}) and the Hierarchy of Pure States (\ac{HOPS}) to \(N\) baths for Hamiltonians of the form of~\cref{eq:generalmodel}.

\subsection{NMQSD}
\label{sec:nmqsd-multi}

Following the usual derivation of the \ac{NMQSD} \cite{Diosi1998Mar}, we
switch to an interaction picture with respect to the \(H_\bath\)
leading to
\begin{equation}
  \label{eq:multimodelint}
  \begin{multlined}[][\bibsep]
      H(t) = H_\sys + \sum_{n=1}^N \qty[L_n^\dag B_n(t) + \hc]\\
     \qq{with} B_n=\sum_{\lambda} g_\lambda\nth a_\lambda\nth\eu^{-\iu \omega_\lambda\nth t}.
  \end{multlined}
\end{equation}
We will discuss the zero temperature case. 
The finite temperature methods generalize straightforwardly to multiple baths. 
Projecting on a Bargmann (non-normalized) coherent state basis \(\qty{\ket{\vb{z}^{(1)},\dots, \vb{z}^{(N)}}= \ket{\underline{\vb{z}}}}\) of the baths yields the following representation of the global state
\begin{equation}
  \label{eq:projected}
  \ket{\Psi(t)} = \int\prod_{n=1}^N{\qty(\frac{\dd{\vb{z}\nth}}{\pi}\eu^{-\abs{\vb{z}}^2})}\ket{\psi(t,\underline{\vb{z}}^\ast)}\ket{\underline{\vb{z}}}.
\end{equation}
Defining the vector process
\begin{equation}
  \label{eq:processes}
  \begin{multlined}
      \pmb{\eta}_t = (\eta^\ast_1(t), \dots, \eta^\ast_N(t)) \\
      \qq{with} \eta^\ast_n(t) = -\iu \sum_\lambda g_\lambda^{\ast (n)} z_\lambda^{\ast (n)}\eu^{\iu \omega_\lambda\nth t}
  \end{multlined}
\end{equation}
and using
\(\pdv{z_\lambda^{\ast(n)}}=\int\dd{s}\pdv{\eta^\ast_n(s)}{z_\lambda^{\ast(n)}}\fdv{\eta^\ast_n(s)}\),
we arrive at
\begin{equation}
  \label{eq:multinmqsd}
  \begin{multlined}[][\bibsep]
      \partial_t\psi_t(\pmb{\eta}^\ast_t) 
  = - \iu H \psi_t(\pmb{\eta}^\ast_t) 
    + \sum_n L_n \eta_n^\ast(t) \psi_t(\pmb{\eta}^\ast_t) 
    \\- \sum_{n=1}^N L_n^\dag\int_0^t\dd{s}\alpha_n(t-s)\fdv{\psi_t(\pmb{\eta}^\ast_t)}{\eta^\ast_n(s)} \,,
  \end{multlined}
\end{equation}
where \(\alpha_n(t-s) = \sum_\lambda |g_\lambda\nth|^2\eu^{-\iu \omega_\lambda\nth(t-s)}\) are the
zero temperature bath correlation functions encoding the structure of the $n$th bath.
Equation \eqref{eq:multinmqsd} becomes \ac{NMQSD} by reinterpreting the \(\vb{z}\nth\) as normal distributed complex random variables by
virtue of Monte-Carlo integration of \cref{eq:projected}. 
The \(\eta^\ast_n(t)\) become Gaussian stochastic processes
defined through
\begin{equation}
  \label{eq:processescorr}
  \begin{multlined}
    \mathcal{M}(\eta^\ast_n(t)) =0, \quad \mathcal{M}(\eta_n(t)\eta_m(s)) = 0, \\
    \mathcal{M}(\eta_n(t)\eta_m(s)^\ast) = \delta_{nm}\alpha_n(t-s).
  \end{multlined}
\end{equation}

\subsection{Nonlinear NMQSD}
\label{sec:nonlin-multi}

For the derivation of the nonlinear theory, the characteristic
trajectories of the partial differential equation of motion of
the bath Husimi function
\begin{equation}
  \label{eq:husimi}
  Q_t(\underline{\vb{z}}, \underline{\vb{z}}^\ast) =
  \frac{\eu^{-\abs{{\underline{\vb{z}}}}^2}}{\pi}
  \braket{\psi(t, {\underline{\vb{z}}})}{\psi(t, {\underline{\vb{z}}}^\ast)}
\end{equation}
have to be determined. 

Using \(\partial_{\underline{\vb{z}}}\ket{\psi(t, {\underline{\vb{z}}}^\ast)} =
0\) and \(\partial_{\underline{\vb{z}}^\ast}\bra{\psi(t, {\underline{\vb{z}}})} =
0\), because \(\ket{\psi(t, {\underline{\vb{z}}}^\ast)}\) is holomorphic in \({\underline{\vb{z}}}^\ast)\),
we derive
\begin{equation}
  \label{eq:husimimotion}
  \partial_tQ_t(\underline{\vb{z}}, \underline{\vb{z}}^\ast) = -i
  \sum_{n=1}^N\qty[\partial_{z_\lambda^{\ast (n)}}\eu^{-\iu \omega_\lambda\nth
    t}\ev{L^\dag_n}_tQ_t(\underline{\vb{z}}, \underline{\vb{z}}^\ast) - \cc],
\end{equation}
where
\begin{equation}
\ev{L^\dag_n}_t = \frac{\mel{\psi(t, {\underline{\vb{z}}})}{L^\dag_n}{\psi(t, {\underline{\vb{z}}}^\ast)}}{\braket{\psi(t, {\underline{\vb{z}}})}{\psi(t, {\underline{\vb{z}}}^\ast)}}.
\end{equation}

The characteristics of \cref{eq:husimimotion} obey the equations of
motion
\begin{equation}
  \label{eq:characteristics}
  \dot{z}^{\ast (n)}_\lambda = \iu g_\lambda\nth \eu^{-\iu \omega_\lambda\nth t} \ev{L^\dag_n}_t
\end{equation}
for the stochastic state labels.

The microscopic dynamics can in-turn be gathered into a shift of the
stochastic processes
\begin{equation}
  \label{eq:procshift}
  \tilde{\eta}_n^\ast(t) = \eta_n^\ast(t) + \int_0^t\dd{s}\alpha_n^\ast(t-s)\ev{L^\dag_n}_s
\end{equation}
and we obtain the nonlinear \ac{NMQSD} equation
\begin{equation}
  \label{eq:multinmqsdnonlin}
  \begin{multlined}
  \partial_t\psi_t(\pmb{\tilde{\eta}}^\ast_t) 
  =  -\iu H \psi_t(\pmb{\tilde{\eta}}^\ast_t) + \sum_n L_n \tilde{\eta}_n^\ast(t)\psi_t(\pmb{\tilde{\eta}}^\ast_t) \\
     - \sum_{n=1}^N \qty(L_n^\dag-\ev{L^\dag_n}_t) \int_0^t\dd{s}\alpha_n(t-s) {\fdv{\psi_t(\pmb{\tilde{\eta}}^\ast_t)}{\eta^\ast_n(s)}}.
  \end{multlined}
\end{equation}
Here the evaluation of the stochastic process at time \(s\) in the
functional derivative is
understood as
\({\eta_{n}^\ast(s) = \eta_{n}(\underline{\vb{z}}^\ast(t), s)}\). This means
that we replace the microscopic \(z_\lambda^{\ast (n)}\) in
\cref{eq:processes} with the shifted ones obeying
\cref{eq:characteristics} and evaluate the resulting function at \(s\).
This awkward construction can be remedied by the convolutionless
formulation. It plays no great role in the \ac{HOPS} formalism.

\subsection{Multi-bath HOPS in Fock space formulation}
\label{sec:multihops}

Following the usual derivation~\cite{RichardDiss} (but with a different unitless normalization) with the exponential expansion of the
BCFs \(\alpha_n(\tau)=\sum_{\mu=1}^{M_n}\,G_\mu\nth\eu^{-W_\mu\nth \tau}\), we define
\begin{equation}
  \label{eq:dops}
  D_\mu\nth(t) \equiv \int_0^t\dd{s}G_\mu\nth\eu^{-W_\mu\nth (t-s)}\fdv{\eta^\ast_n(s)}
\end{equation}
and
\begin{equation}
  \label{eq:dops_full}
  D^{\underline{\vb{k}}} \equiv
  \prod_{n=1}^N\prod_{\mu=1}^{M_n}
  {\sqrt{\frac{1}{\qty(G\nth{}_\mu)^{\underline{\vb{k}}_{n,\mu}}\underline{\vb{k}}_{n,\mu}!}}
  \frac{1}{\iu^{\underline{\vb{k}}_{n,\mu}}}}\qty(D_\mu\nth{}(t))^{\underline{\vb{k}}_{n,\mu}},
\end{equation}
as well as
\begin{equation}
  \label{eq:hierdef}
  \psi^{\underline{\vb{k}}} \equiv D^{\underline{\vb{k}}}(t)\psi.
\end{equation}
Here $\kmat$ denotes the multi-index vector $(\vb{k_1}, \dots \vb{k_N})$.
Using
\begin{equation}
  \label{eq:commrelation}
  [D^\kmat(t),\eta_n^\ast(t)] =  -\iu\sum_{\mu=1}^{M_n}
  \sqrt{\kmat_{n,\mu}G\nth_\mu} D^{\kmat -
    \mat{e}_{n,\mu}}
\end{equation}
where \({\qty(\mat{e}_{n,\mu})}_{ij}=\delta_{ni}\delta_{\mu j}\) we find after some
algebra
\begin{widetext}
\begin{equation}
\label{eq:multihops}
    \dot{\psi}^\kmat
    = \qty(-\iu H_\sys + \sum_n L_n \cdot\eta_n^\ast(t) - \sum_{n=1}^N\sum_{\mu=1}^{M_n}\kmat_{n,\mu}W\nth_\mu)\psi^\kmat 
       - \iu \sum_{n=1}^N\sum_{\mu=1}^{M_n}\sqrt{G\nth_\mu}\qty(\sqrt{\kmat_{n,\mu}}  L_n\psi^{\kmat - \mat{e}_{n,\mu}} 
      + \sqrt{\qty(\kmat_{n,\mu} + 1)}  L^\dag_n\psi^{\kmat + \mat{e}_{n,\mu}}).
\end{equation}
\end{widetext}

The \ac{HOPS} equations \eqref{eq:multihops} can also be rewritten in an
especially appealing form~\cite{Gao2021Sep} if we embed the hierarchy
states into a larger Hilbert space using
\begin{equation}
  \label{eq:fockpsi}
  \ket{\Psi} = \sum_\kmat\ket{\psi^\kmat}\otimes \ket{\kmat}
\end{equation}
where
\(\ket{\kmat}=\bigotimes_{n=1}^N\bigotimes_{\mu=1}^{N_n}\ket{\kmat_{n,\mu}}\)
are bosonic Fock states.
Now \cref{eq:multihops} becomes
\begin{equation}
  \label{eq:fockhops}
  \begin{aligned}
    \partial_t\ket{\Psi} ={}& \begin{aligned}[t]
                      \Bigg( 
                      & -\iu H_\sys + \sum_n L_n \eta_n^\ast(t) 
                        - \sum_{n=1}^N\sum_{\mu=1}^{M_n}b_{n,\mu}^\dag b_{n,\mu} W\nth_\mu \\
                      & - \iu \sum_{n=1}^N\sum_{\mu=1}^{M_n} \sqrt{G_{n,\mu}} \qty(b^\dag_{n,\mu}L_n + b_{n,\mu}L^\dag_n) \Bigg) \ket{\Psi}
                    \end{aligned}
    \\ \equiv{}& -\iu H_\mathrm{eff}\ket{\Psi}   
  \end{aligned}    
\end{equation}
defining the effective non-hermitian stochastic Hamiltonian $H_\mathrm{eff}$.

The non-linear generalization of \cref{eq:fockhops} is
\begin{equation}
  \label{eq:fockhopsnonlin}
  \begin{multlined}
  \partial_t\ket{\Psi} = \Big( 
   -\iu H_\sys + \sum_n L_n \tilde{\eta}_n^\ast(t) 
    - \sum_{n=1}^N\sum_{\mu=1}^{M_n}b_{n,\mu}^\dag b_{n,\mu} W\nth_\mu \\
   - \iu \sum_{n=1}^N\sum_{\mu=1}^{M_n} \sqrt{G_{n,\mu}} \qty[b^\dag_{n,\mu}L_n + b_{n,\mu}\pqty{L^\dag_n-\ev{L^\dag_{n}}_{t}}] \Big) \ket{\Psi},
  \end{multlined}
\end{equation}
where \(\tilde{\eta}\) is the shifted stochastic process~\cref{eq:shifted_proc}.

\section{General collective bath observables}
\label{sec:gener-coll-bath}

In the main text, we have introduced the formalism for bath observables using the example of the
bath energy flow \(J\). Here, we proceed to more
general observables of the form
\begin{equation}
  \label{eq:collective_obs}
  O = f(B^\dag, B) = \sum_{\alpha}F_\alpha\otimes \qty(B_{i}^\dag)^{\alpha_1}\pqty{B_{j}}^{\alpha_2}
\end{equation}
where \(\alpha\) is a two-dimensional multi-index, \(B\) is as
in~\cref{eq:totalH}, \(i,j\) index the bath and the \(F_\alpha\) are
general observables acting on the system only.
We will restrict
the discussion to the case of a single bath, as the generalization to
multiple baths is straightforward.

To evaluate \(\ev{O}\), we have to find the value of
\(\ev{\qty(B^\dag)^a B^b}\). This can be achieved by interjecting the
overcomplete coherent state basis as in \cref{eq:interactev} which
leads us to expressions in the form of
\(\mel{\psi}{\qty(B^\dag)^a}{z}\mel{z}{B^b}{\psi}\) and switching to the
interaction picture with respect to \(H_{\bath}\).

For zero temperature, we find, following the procedures of
\cref{sec:bath-energy-change} we find
\begin{equation}
  \label{eq:bmel}
  \begin{aligned}
      \mel{z}{B^b(t)}{\psi} 
    &{}= (-\iu D(t))^b\ket{\psi(\eta^\ast,t)}\\
    &{}= (-\iu)^b\sum_{\abs{\vb{k}}=b}\binom{b}{\vb{k}} \iu^{\vb{k}} \sqrt{{G^{\vb{k}}}{\vb{k}!}}\ket{\psi^{\vb{k}}}.
  \end{aligned}
\end{equation}

By the same means we obtain
\begin{equation}
  \label{eq:bdagmel}
  \begin{aligned}
      \mel{\psi}{\qty(B^\dag(t))^a}{z} 
   &{}= \qty(\mel{z}{B^a}{\psi})^\dag
   = \qty((-\iu D(t))^a\ket{\psi(\eta^\ast,t)})^\dag \\
   &{}= (\iu)^a\sum_{\abs{\vb{k}}=a}\binom{a}{\vb{k}} (-\iu)^{\vb{k}} \sqrt{{\qty(G^{\vb{k}})^\ast}{\vb{k}!}}\bra{\psi^{\vb{k}}}.   
  \end{aligned}
\end{equation}

In the above \(\vb{k}! = k_1!k_2!\ldots\) and
\(G^{\vb{k}}=G_1^{k_1}G_2^{k_2}\ldots\) following the usual conventions for
multi-indices. Thus, expressions involving the bath operator \(B\) to
the \(b\)th power lead to expressions involving the hierarchy states
of level \(b\). 

Returning to \cref{eq:collective_obs}, we find
\begin{widetext}
  \begin{equation}
  \label{eq:f_ex_zero}
  \ev{O} = \mathcal{M}_{\eta^\ast}\sum_{\alpha} \sum_{\substack{\abs{\gamma}=\alpha_1\\\abs{\delta}=\alpha_2}}
  \Bigg[\binom{\alpha_1}{\gamma}\binom{\alpha_2}{\delta}(\iu)^{\delta+\alpha_1}
  \times (-\iu)^{\gamma+\alpha_2}
    \sqrt{{\qty(G^\gamma)^\ast G^\delta}{\gamma!\delta!}}
    \mel{\psi^\gamma}{F_\alpha}{\psi^\delta}\Bigg].
\end{equation}
\end{widetext}

For finite temperatures we substitute \(B(t)\to B(t)+\xi(t)\)
(interaction picture) as in
\cref{sec:non-zero_temperature} to obtain
\begin{equation}
  \label{eq:finite_arb}
  \begin{multlined}[][\columnwidth]
      \mel{z}{\qty(B(t)+\xi(t))^b}{\psi} \\= \sum_{m=0}^b \binom{b}{m} \xi^{b-m}(t)(-\iu D(t))^m\ket{\psi(\eta^\ast,t)}
  \end{multlined}
\end{equation}
which may be substituted into the above.

The nonlinear method can be accommodated as in \cref{sec:nmqsd}. For the expressions
like~\cref{eq:f_ex_zero} involving the \ac{HOPS} hierarchy states, the
method can be implemented by dividing by the squared norm of the
zeroth hierarchy state.

The generalization to multiple baths may be performed in the same
manner as was discussed in \cref{sec:hops_multibath}. This allows us to
calculate the expectation value involving multiple bath operators
\(B^{(n)}\). Interestingly, the generalization of \cref{eq:bmel} to
multiple baths immediately links hierarchy states of the form
\(\ket{\psi^{\underline{e}_{i,k_i} + \underline{e}_{j,k_j}}}\) (see
\cref{sec:multihops} for notation) with correlations between the baths
\(\ev{B_i(t)B_j(t)}\). Under the assumption that these correlations
remain small, one could argue to neglect them when performing
numerical calculations. Verifying this behavior would be an
interesting task for future work.

By anti-normal ordering the \(B, B^\dag\) in~\cref{eq:collective_obs}
and inserting the coherent state resolution of unity we find terms of
the form
\begin{equation}
  \label{eq:with_process}
  \mel{z}{\qty(B^\dag(t))^b}{\psi} \sim \qty(\eta^\ast_{t})^b\ket{\psi(\eta^\ast,t)}.
\end{equation}
The corresponding version of~\cref{eq:f_ex_zero} then only
depends on the zeroth order state and the stochastic processes. It has
been observed that expressions involving the stochastic process
directly tend to converge more slowly. However, this statement comes
without proof and its verification may be left to future study. An
explanation may be that the first hierarchy states fluctuate about
their average dynamics, whereas the stochastic process fluctuates
around zero and does not contain much information about the actual
dynamics.

If one wants to obtain information about \emph{part} of the bath
\cref{eq:with_process} could become useful. The bath could be
decomposed into several sub-baths by virtue of expressing the spectral
density as a sum of functions with disjoint support, leading to
multiple stochastic processes which would add up to the original
process. In this way, the hierarchy structure could remain unaltered,
while \cref{eq:with_process} could be evaluated for the sub-processes.

Also, this alternative method could be used as a convergence and
consistency check, as expressions of the form~\cref{eq:with_process}
involve the hierarchy cutoff and the exponential expansion of the \ac{BCF}
only in an indirect manner.

\section{Notes on the numerics}
\label{sec:notes-numerics}

Most simulations used \(N={8\cdot 10^{4}}\) samples, a hierarchy depth of \(k_{\max} = 4\) and a \ac{BCF} expansion with five terms for each bath. The relative error in energy conservation over the three cycles simulated was usually less than \(1\%\text{--}2\%\) and less than \(1\%\) in the limit cycle. The energy conservation was assessed by comparing the sum of integrated bath energy change, system energy and interaction energy against the integrated total power.

Our back-end code implementing the \ac{HOPS} is available at \href{https://github.com/OpQuSyD/hops}{github.com/OpQuSyD/hops}.
Further available code snippets related to this work are 
the \href{https://github.com/vale981/HOPSFlow-Paper}{front-end code}\footnote{\url{https://github.com/vale981/HOPSFlow-Paper}}, 
the implementation to access \href{https://github.com/vale981/HOPSFlow}{bath related observables}\footnote{\url{https://github.com/vale981/HOPSFlow}} and 
the \href{https://github.com/vale981/two_qubit_model/blob/main/hiro_models/otto_cycle.py}{Otto cycle configuration}\footnote{\url{https://github.com/vale981/two_qubit_model/blob/main/hiro_models/otto_cycle.py}} for the HOPS algorithm. 

Note also that an alternative implementation of the \ac{HOPS} with an adaptive hierarchical structure capable of size-invariant scaling for large molecular aggregates~\cite{varvelo_formally_2021} can be found at
\href{https://github.com/MesoscienceLab/mesohops}{github.com/MesoscienceLab/mesohops}.

\end{document}